\documentclass{PoS}
\usepackage[utf8]{inputenc}
\usepackage{amsmath,amssymb,bm}
\usepackage{graphicx}
\usepackage{epsfig}
\usepackage{braket}
\usepackage{slashed}
\usepackage{multirow}
\usepackage{float}
\usepackage{rotating}

\usepackage{xcolor}
\definecolor{Darkgreen}{RGB}{30,120,30}

\def\gsim{\mathrel{\rlap{\lower4pt\hbox{\hskip1pt$\sim$}}
    \raise1pt\hbox{$>$}}}         
\def\lsim{\mathrel{\rlap{\lower4pt\hbox{\hskip1pt$\sim$}}
    \raise1pt\hbox{$<$}}}         

\newcommand{\be}{\begin{equation}}
\newcommand{\ee}{\end{equation}}
\newcommand{\bea}{\begin{eqnarray}}
\newcommand{\eea}{\end{eqnarray}}
\newcommand{\bi}{\begin{itemize}}
\newcommand{\ei}{\end{itemize}}
\newcommand{\ben}{\begin{enumerate}}
\newcommand{\een}{\end{enumerate}}

\newcommand{\lp}{\left(}
\newcommand{\rp}{\right)}
\newcommand{\as}{\alpha_s}

\newcommand{\gA}{A}
\newcommand{\oP}{P} 
\def\good{\makebox[1em]{\centering{\mbox{\color{green}$\bigstar$}}}}
\def\bad{\makebox[1em]{\centering\color{red}\tiny$\blacksquare$}}
\def\soso{\makebox[1em]{\centering{\mbox{\raisebox{-0.5mm}{\color{green}\Large$\circ$}}}}}

\def\gbar{\bar{g}}
\newcommand{\msbar}{{\overline{{\rm MS}}}}
\newcommand{\Nf}{N_{\hspace{-0.08 em} f}}
\newcommand{\nl}{n_{\hspace{-0.08 em} \rm l}}
\newcommand{\rmO}{\mathcal{O}}
\newcommand{\Refs}{\,\mathrm{Refs.}}

\newcommand{\GeV}{\,\mathrm{GeV}}
\newcommand{\MeV}{\,\mathrm{MeV}}
\newcommand{\fm}{\,\mathrm{fm}}
\def\oO{\mathcal{Q}}
\newcommand{\simas}[1]{\raisebox{-.1ex}{
            $\stackrel{\small{#1}}{\sim}$}}

\newcommand{\eq}[1]{Eq.~(\ref{#1})}
\newcommand{\fig}[1]{Fig.~\ref{#1}}
\newcommand{\tab}[1]{Tab.~\ref{#1}}
\newcommand{\sect}[1]{Sec.~\ref{#1}}

\title{Determining the strong coupling: status and challenges}

\ShortTitle{The strong coupling: status and challenges}

\author{Antonio Pich\\
IFIC, Universitat de Val\`encia -- CSIC, Apt. Correus 22085, E-46071 Val\`encia, Spain.
\\
E-mail: \email{pich@ific.uv.es}}

\author{Juan Rojo\\
Department of Physics and Astronomy, VU University, De Boelelaan 1081,
1081 HV Amsterdam,\\ 
and Nikhef Theory Group, Science Park 105, 1098 XG Amsterdam, The Netherlands. 
\\
E-mail: \email{j.rojo@vu.nl}}

\author{Rainer Sommer\\
John von Neumann Institute for Computing (NIC), DESY, Platanenallee~6, D-15738~Zeuthen, Germany. \\
E-mail: \email{rainer.sommer@desy.de}}

\author{Antonio Vairo\\
Physik-Department, Technische Universit\"at M\"unchen, 
James-Franck-Str. 1, 85748 Garching, Germany. \\
E-mail: \email{antonio.vairo@tum.de}}

\abstract{
  The ``XIIIth Quark Confinement and the Hadron Spectrum''
  conference (Confinement 2018) contained a ``Round Table Discussion''
    on the status of the
  determinations of the strong coupling $\alpha_s(m_Z)$
  as well as prospects for future improvements.
  In this contribution, we summarize the different aspects
  of the discussion. In particular, we cover $\as$ determinations
  from inclusive observables, such as the hadronic decays
  of the $Z$ boson and the $\tau$ lepton;
  from global fits of parton distribution functions (PDFs);
  from high-energy collider observables, and from event shapes;
  as well as from various observables computed by lattice QCD,
  specifically from the QCD static energy.
  There is overall good agreement between these various
  determinations, but there are also outliers,
  differing from the world average by up to $-5\%$. Nevertheless, the 
  general agreement constitutes a beautiful and 
  significant test of the detailed nature 
  of the strong interactions, and provides a crucial input
  for high-precision calculations in QCD and beyond.
}

\FullConference{XIII Quark Confinement and the Hadron Spectrum - Confinement2018\\
		31 July - 6 August 2018\\
		Maynooth University, Ireland}

\begin{document}

\section{Introduction}
Quantum Chromodynamics (QCD) contains only a single free
parameter, the value of its coupling constant, besides the values
of the quark masses. Therefore, all strong interaction phenomena
should be described in terms of the unique strong coupling
$\alpha_s$. The overwhelming consistency of the many determinations of
$\alpha_s$, performed in very different processes and in a broad range
of energy scales, provides a beautiful verification of QCD,
establishing the SU(3)${}_C$ gauge theory as the fundamental quantum
field theory of the strong interaction.

The impressive progress achieved in recent years has promoted
perturbative QCD to the status of precision physics, improving in a
very sizeable way the physics potential of the current LHC
program.
In the $\msbar$-scheme
the QCD $\beta$ function is 
known to five loops \cite{Baikov:2016tgj,Luthe:2016ima,Herzog:2017ohr,Luthe:2017ttg}, which provides a
precise theoretical control of the renormalization-scale dependence of
the running coupling. Moreover, the very modest growth of the
$\beta$-function coefficients with the perturbative order gives rise
to a surprisingly smooth power expansion. The $\beta$ function and,
therefore, the running coupling depend on the number of active quark
flavours $n_f$. The matching conditions relating the effective QCD
theories with $n_f$ and $n_f-1$ quark flavours are known to four loops
\cite{Schroder:2005hy,Chetyrkin:2005ia}.

At the current level of accuracy, a very good understanding of the
uncertainties associated with the different measurements of $\alpha_s$
is needed. Thus, we should restrict the analysis to observables where
perturbative techniques are reliable and enough terms in the expansion
in powers of $\alpha_s$ are available. Following
the Particle Data Group (PDG) criteria~\cite{Tanabashi:2018oca},
in this contribution we will require a
NNLO or higher accuracy. Non-perturbative contributions and
theoretical uncertainties from the expected asymptotic behaviour of
perturbation theory should also be under good control.

The outline of this contribution is the following.
  First of all, in Sect.~\ref{sec:inclusive}, we review the
  determinations of $\alpha_s$ from inclusive
  observables such as the hadronic $\tau$ decay width and the
  $Z\to{\rm hadrons}$ branching fraction.
  Then, in Sect.~\ref{sec:pdffits}, we discuss determinations
  from global PDF fits, high-energy collider observables,
  and event shapes.
  In Sect.~\ref{sec:static}, we study extractions of the strong
  coupling from the lattice QCD static energy.
  Then in Sect~\ref{sec:lattice}, we review recent progress
  in determining $\alpha_s(m_Z)$ from lattice QCD calculations
  more generally.
  Finally in Sect.~\ref{sec:finaldiscussion} we present
  some general reflections about the main lessons that can be drawn
  from this discussion exercise.

\section{Inclusive observables \hskip .2cm {\it (A. Pich)}}
\label{sec:inclusive}

The inclusive electroweak production of hadrons provides clean
observables that can be accurately predicted with perturbative tools,
such as the cross section $\sigma(e^+e^-\to\mathrm{hadrons})$ or the
decay widths $\Gamma(Z\to\mathrm{hadrons})$ and
$\Gamma(W\to\mathrm{hadrons})$. The final hadrons are produced through
the colour-singlet vector $\, V^{\mu}_{ij} = \bar{\psi}_j \gamma^{\mu}\psi_i \, $ and axial-vector $\, A^{\mu}_{ij} = \bar{\psi}_j
\gamma^{\mu} \gamma_5 \psi_i \,$ quark currents ($i,j=u,d,s\ldots$). 
The QCD dynamics is governed by the corresponding two-point correlation functions ($J=V,A$)
\begin{equation}\label{eq:pi_v2}
\Pi^{\mu \nu}_{ij,J}(q)\; \equiv\;
 i \int d^4x \;\, \mathrm{e}^{iqx}\,
\langle 0|T(J^{\mu}_{ij}(x)\, J^{\nu}_{ij}(0)^\dagger)|0\rangle
\; =\;
\left( -g^{\mu\nu} q^2 + q^{\mu} q^{\nu}\right) \, \Pi_{ij,J}^{(0+1)}(q^2)
 +   g^{\mu\nu} q^2\, \Pi_{ij,J}^{(0)}(q^2) \, .
\end{equation}
The scalar functions $\Pi_{ij,J}^{(L)}(q^2)$ 
(the superscript $L=0,1$ denotes the angular momentum in the hadronic rest frame)
are analytic in the whole complex $q^2$ plane, except along the (physical)
positive real axis where their imaginary parts have discontinuities. 
These absorptive cuts correspond to the measurable
hadronic spectral distributions with the given quantum numbers.

For massless quarks, $\Pi_{ij,V}^{(0)}(s)= 0$ while $s\,
\Pi_{ij,A}^{(0)}(s)$ is a known constant generated by the
non-perturbative Goldstone-pole contribution that cancels in
$\Pi_{ij,A}^{(0+1)}(s)$.  When $i\not= j$, the two quark currents must
necessarily be connected through a quark loop (non-singlet topology),
which gives identical contributions to the vector and axial massless
correlators:\ $\Pi_{i\not=j,V}^{(0+1)}(s)=\Pi_{i\not=j,A}^{(0+1)}(s)$.
The neutral-current correlators ($i=j$)
get additional singlet contributions, where each current couples to a
different quark loop. Since gluons have $J^{PC}=1^{--}$ and colour,
these singlet topologies start to contribute at
$\mathcal{O}(\alpha_s^3)$ and $\mathcal{O}(\alpha_s^2)$, respectively,
for the vector and axial-vector currents.

The perturbative expansion of the correlators $\Pi_{ij,J}^{(0+1)}(s)$ is known with an impressive $\mathcal{O}(\alpha_s^4)$ accuracy \cite{Baikov:2008jh,Baikov:2012er,Baikov:2012zn}.
Therefore, the ratio
$R_{e^+e^-}(s)\equiv\sigma(e^+e^-\to
\mathrm{hadrons})/\sigma(e^+e^-\to\mu^+\mu^-)$, 
which is proportional to the imaginary part of the electromagnetic vector-current
correlator, could be used to perform a clean
N${}^3$LO determination of $\alpha_s$, at energies high enough to safely neglect non-perturbative contributions. The experimental uncertainties are, however, too large to get a competitive value.

\paragraph{$\boldsymbol{\Gamma(Z\to\mathrm{\bf hadrons})}$.}
\label{subsec:Zwidth}

The electroweak neutral quark current contains vector and axial-vector
components, weighted with the corresponding $Z$ couplings. The large
value of the top mass generates sizeable singlet axial corrections
which start at $\mathcal{O}(\alpha_s^2)$. The ratio of the hadronic
and electronic widths of the $Z$ boson is given by the perturbative
QCD series ($m_b=0$, $m_t\not=0$, $n_f=5$)
\begin{equation}\label{eq:RZ}
R_Z\; \equiv\; \frac{\Gamma(Z\to\mathrm{hadrons})}{\Gamma(Z\to e^+e^-)}\; =\; R_Z^{\mathrm{EW}} \; N_C\;\left\{
1 + \sum_{n=1}\; \tilde F_n
\left( {\alpha_s(m_Z)\over \pi}\right)^n\right\}
\, ,
\end{equation}
with $\tilde F_1 = 1$, $\tilde F_2 = 0.76264$,
$\tilde F_3 = -15.490$ and $\tilde F_4 = -68.241$ \cite{Baikov:2012er}.
Taking into account electroweak corrections and QCD contributions suppressed by powers of $m_b^2/m_Z^2$ \cite{Chetyrkin:1994js,Chetyrkin:2000zk}, the ratio $R_Z$ is included in the global fit to electroweak precision data. One obtains in this way a quite accurate value of $\alpha_s(m_Z)$ \cite{Baak:2014ora}:
\begin{equation}\label{eq:alpha_Z}
\alpha_s^{(n_f=5)}(m_Z)
\; =\; 0.1196\pm 0.0030\, .
\end{equation}
Since this result assumes the validity of the electroweak Standard Model, its comparison with other determinations of the strong coupling provides a non-trivial constraint on new-physics scenarios.

\paragraph{$\boldsymbol{\Gamma(\tau\to\nu_\tau + \mathrm{\bf hadrons})}$.}
\label{subsec:Tauwidth}

\begin{figure}[t]
\centerline{
\includegraphics[width=0.36\textwidth]{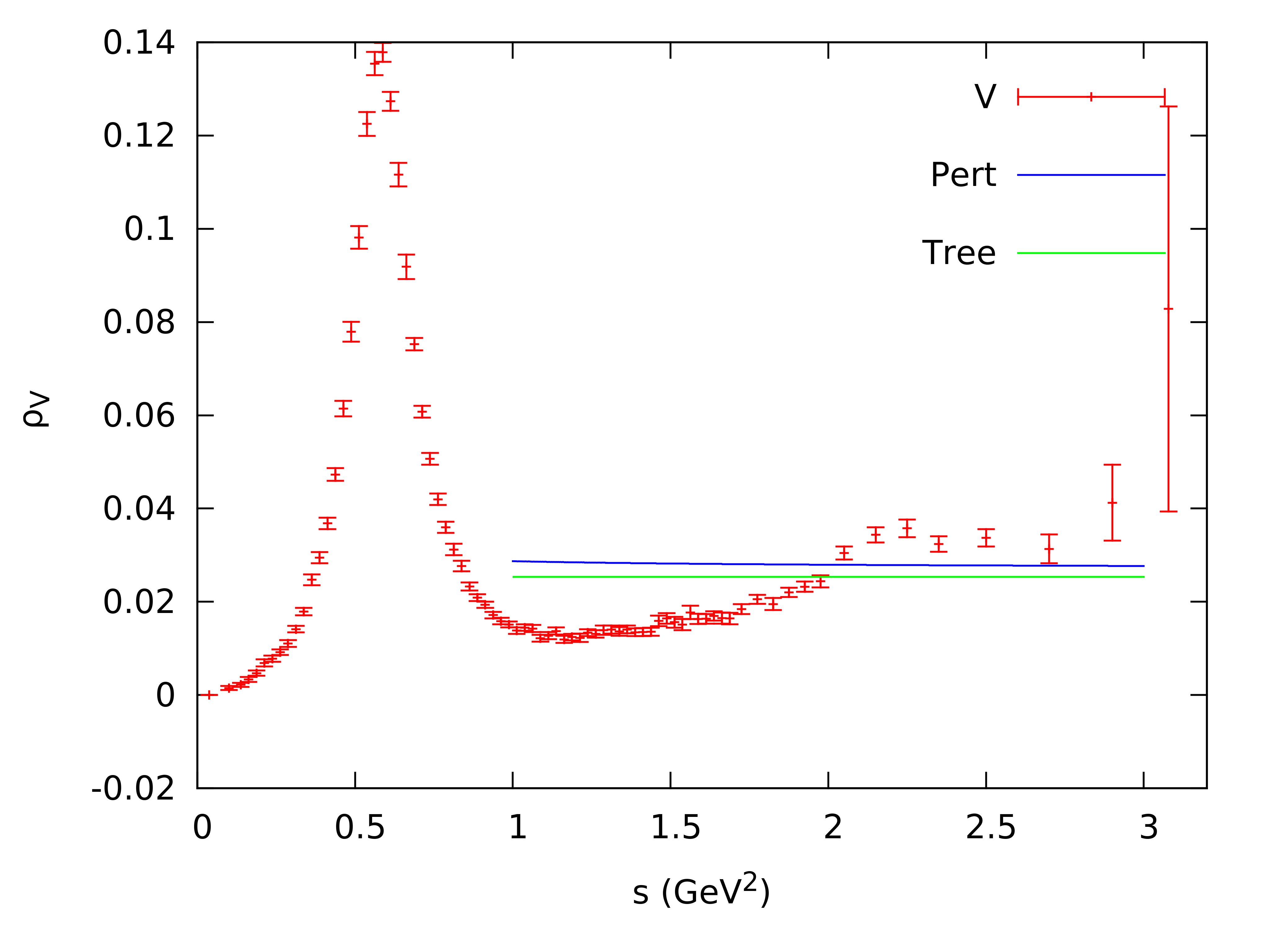}\hskip -.15cm
\includegraphics[width=0.36\textwidth]{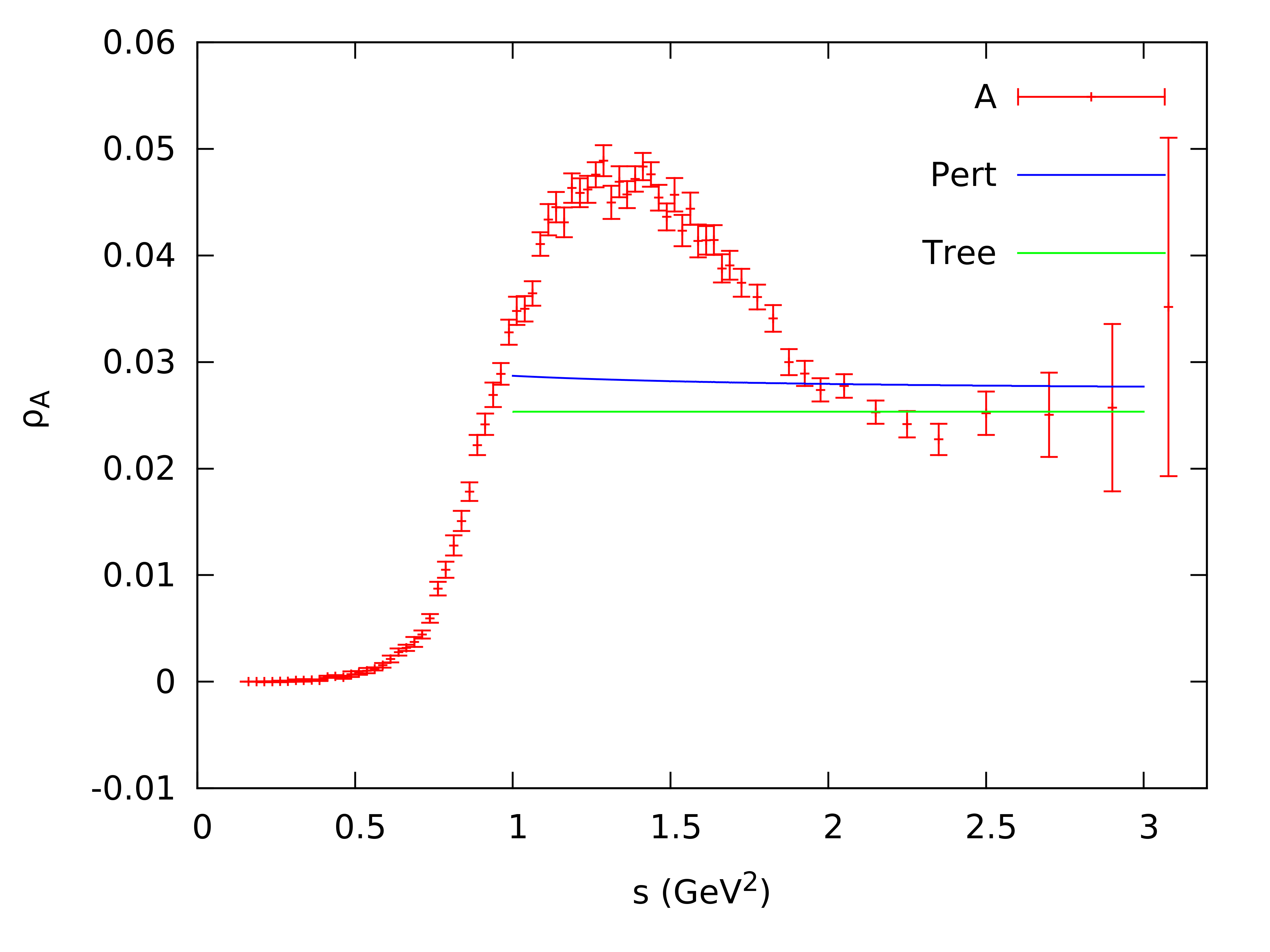}\hskip -.15cm
\includegraphics[width=0.36\textwidth]{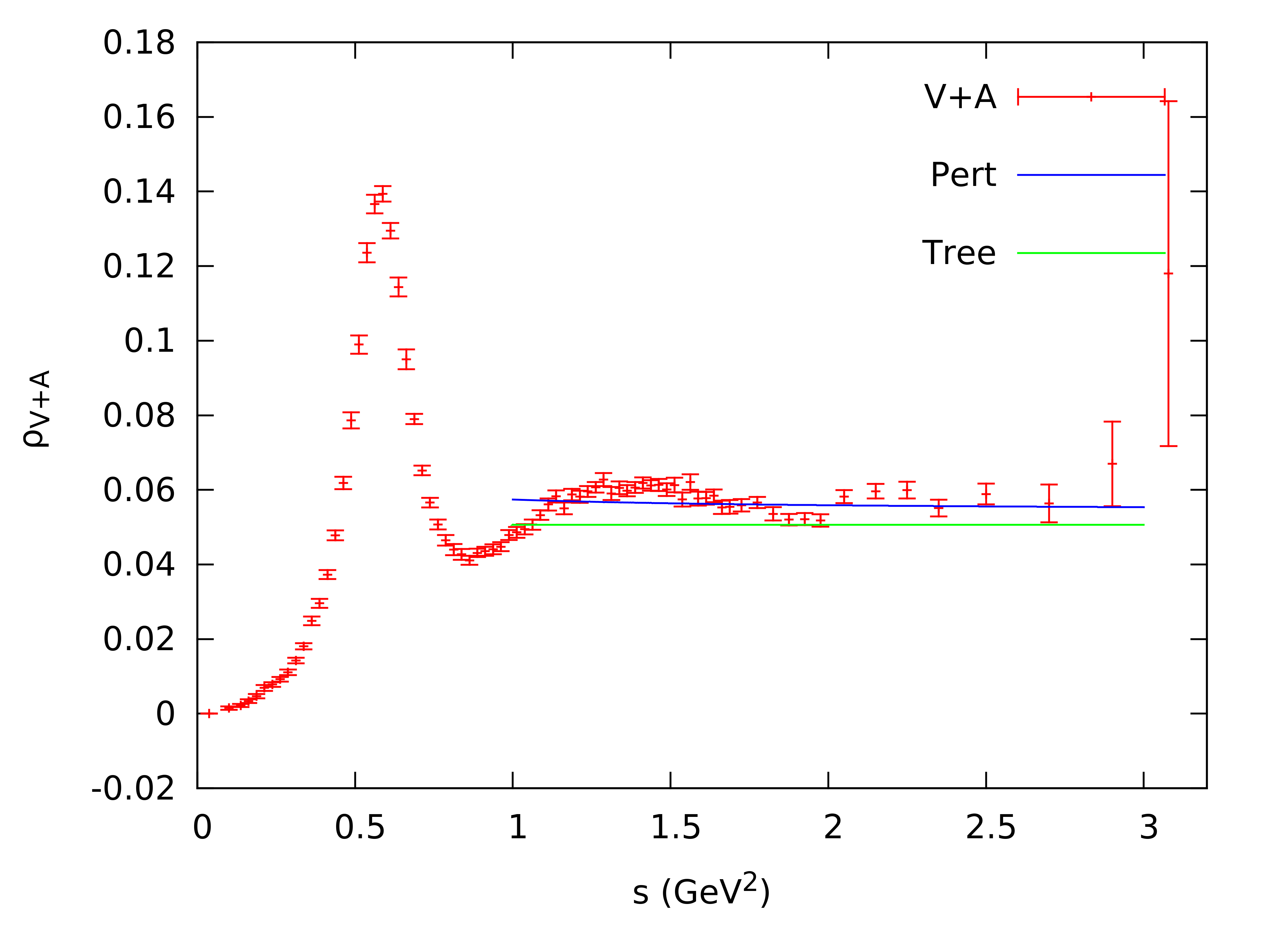}
}
\caption{Spectral functions for the $V$, $A$ and $V+A$ channels, determined from ALEPH $\tau$ data~\cite{Davier:2013sfa}.}
\label{fig:SpectralFunction}
\end{figure}

The current precision on the hadronic decay width of the $W^\pm$ boson is not good enough to perform an accurate determination of $\alpha_s$. A much more sensitive observable is the hadronic $\tau$ decay width \cite{Narison:1988ni,Braaten:1988hc,Braaten:1988ea,Braaten:1991qm}, which proceeds through a virtual $W^\pm$ boson. 
Restricting the analysis to the dominant Cabibbo-allowed decay width,
\begin{eqnarray}\label{eq:R_tau}
R_{\tau,V+A} &\!\!\equiv &\!\!\frac{ \Gamma [\tau^- \to \nu_\tau +\mathrm{hadrons}\, (S=0)]}{ \Gamma [\tau^- \to \nu_\tau e^- {\bar \nu}_e]}
\\ \hskip -.5cm\mbox{}
&\hskip -.5cm \!\! = &\!\! 12 \pi\, |V_{ud}|^2\, S_{\mathrm{EW}} \int^{m_\tau^2}_0 {ds \over m_\tau^2 } \,
 \left(1-{s \over m_\tau^2}\right)^2
\biggl[ \left(1 + 2 {s \over m_\tau^2}\right)
 \mbox{\rm Im} \Pi^{(0+1)}_{ud,V+A}(s)
 - 2 {s \over m_\tau^2}\, \mbox{\rm Im} \Pi^{(0)}_{ud,V+A}(s) \biggr]\,  ,
\nonumber\end{eqnarray}
where $S_{\mathrm{EW}}=1.0201\pm 0.0003$ incorporates the
electroweak radiative corrections \cite{Marciano:1988vm,Braaten:1990ef,Erler:2002mv}.
The measured invariant-mass distribution of the final hadrons determines the
spectral functions $\rho_J(s) \equiv \mathrm{Im} \Pi_{ud,J}^{(0+1)}(s)/\pi$, 
shown in figure~\ref{fig:SpectralFunction} (the only relevant contribution to the $s\,\mathrm{Im} \Pi^{(0)}_{ud,V+A}(s)$ term is the $\pi^-$ final state at $s=m_\pi^2$).

Although the low-energy spectral functions themselves cannot be described with perturbative tools, the analyticity properties of the $\Pi^{(L)}_{ij,J}(s)$ correlators relate weighted integrals (moments) of the experimental spectral distribution with theoretical QCD predictions \cite{Braaten:1991qm,LeDiberder:1992zhd}:
\begin{equation}\label{aomega}
A^{\omega}_J(s_{0})\;\equiv\; \int^{s_{0}}_{s_{\mathrm{th}}} \frac{ds}{s_{0}}\;\omega(s)\; \mathrm{Im} \Pi_{ud,J}^{(0+1)}(s)\; =\; \frac{i}{2}\;\oint_{|s|=s_{0}}
\frac{ds}{s_{0}}\;\omega(s)\, \Pi_{ud,J}^{(0+1)}(s)\, .
\end{equation}
The complex integral in the right-hand side (rhs) runs
counter-clockwise around the circle $|s|=s_{0}$, $s_{\mathrm{th}}$ is
the hadronic mass-squared threshold and $\omega(s)$ is any weight
function analytic in $|s|\le s_0$.  While the left-hand-side integral
is directly determined by the experimental data, for large-enough
values of $s_0$ the operator product expansion (OPE),
$\Pi_{ud,J}^{(0+1)}(s)^{\mathrm{OPE}} = \sum_{D} \mathcal{O}_{D,\,
J}/(-s)^{D/2}$, can be used to predict the contour integral as an
expansion in inverse powers of $s_0$, the perturbative contribution
being the leading $D=0$ term. Contributions to the rhs integral from
the region near the real axis, where the OPE is not valid, can be
efficiently suppressed with `pinched' weights that vanish at $s=s_0$.

The ratio $R_{\tau,V+A}$ involves the doubly-pinched weight $\omega(x)= (1-x^2)(1+2x)=1-3x^2+2x^3$, with $x\equiv s/s_0$ 
and $s_0=m_\tau^2$. Cauchy's theorem implies that the contour integral is
only sensitive to OPE corrections with $D=6$ and 8, which are strongly
suppressed by the corresponding powers of the $\tau$ mass. In
addition, the $D=6$ vector and axial-vector contributions have
opposite signs, cancelling to a large extent. Therefore,
$R_{\tau,V+A}$ is a clean observable
to determine $\alpha_s$. It is known with $\mathcal{O}(\alpha_s^4)$
precision, and it is very sensitive to the strong coupling because
$\alpha_s(m_\tau)$ is sizeable. The small non-perturbative corrections
can be directly estimated from the data, using weights with the
appropriate power of $s$ to project a particular OPE contribution
\cite{LeDiberder:1992zhd}.
The dominant theoretical uncertainty is the perturbative error
associated with the unknown higher-order
corrections.
For a given value of
$\alpha_s$, a truncated fixed-order perturbation theory (FOPT)
approximation \cite{Braaten:1991qm} leads to a larger perturbative
contribution than the so-called contour-improved perturbation theory
(CIPT) \cite{LeDiberder:1992jjr,Pivovarov:1991rh}, which resums large
corrections arising from the long running of the strong coupling along
the circle $s=s_0$. Therefore, FOPT results in a smaller fitted value
of $\alpha_s(m_\tau)$ than~CIPT.

The predicted suppression of non-perturbative corrections
\cite{Braaten:1991qm} has been confirmed through detailed analyses of
the invariant-mass distribution of the hadronic $\tau$ decay products,
performed by ALEPH
\cite{Buskulic:1993sv,Barate:1998uf,Schael:2005am,Davier:2005xq,Davier:2008sk},
CLEO \cite{Coan:1995nk} and OPAL \cite{Ackerstaff:1998yj}, showing
that these effects are below 1\% in $R_{\tau,V+A}$. In comparison, the
purely perturbative contribution is around 20\%. The most complete and
precise experimental study, performed with the recently updated ALEPH
data, gives $\alpha_s^{(n_f=3)}(m_\tau) = 0.332\pm
0.005_{\mathrm{exp}}\pm 0.011_{\mathrm{th}}$ \cite{Davier:2013sfa},
where the second error takes into account the different central values
obtained with the FOPT (0.324) and CIPT (0.341) prescriptions. Taking
as input the small non-perturbative contribution extracted from the
ALEPH analysis, the strong coupling can 
also be determined
from the total $\tau$ hadronic width (and/or lifetime); this gives
$\alpha_s^{(n_f=3)}(m_\tau) = 0.331\pm 0.013$ (FOPT + CIPT)
\cite{Pich:2013lsa}, in perfect agreement with the ALEPH result.

Slightly smaller (10\%) values of $\alpha_s(m_\tau)$ have been
obtained in Ref.~\cite{Boito:2014sta} through a direct fit of the
vector spectral function from $s= \hat s_0 = 1.55\: \mathrm{GeV}^2$ to
$m_\tau^2$, with a 4-parameter ansatz for $\rho_V(s)$, plus the moment
$A_V^{\omega=1}(\hat s_0)$. This approach maximizes the role of
non-perturbative effects in order to better study them \cite{Boito:2016oam,Boito:2018yml}, but this is not a good strategy to perform an accurate determination of the strong coupling. The OPE is actually not valid on the physical cut and the
quoted uncertainties are largely underestimated \cite{Pich:2016mgv,Pich:2018jiy}. A more careful numerical study has shown that the fitted results strongly depend on the chosen value of $\hat s_0$ and the particular form of the assumed
spectral-function ansatz \cite{Pich:2016bdg}: fluctuations of
$\alpha_s(m_\tau)$ larger than $3\sigma$ are obtained with slight
modifications of these assumptions, showing that the fitted value is
model dependent.

\begin{table}[t]
\centering
\begin{tabular}{|c|c|c|c|}
\hline  &\multicolumn{3}{c|}{} \\[-12pt]
Method  & \multicolumn{3}{c|}{$\alpha_{s}^{(n_f=3)}(m_{\tau})$}
\\[1.4pt] \cline{2-4}
& \raisebox{-2pt}{CIPT} & \raisebox{-2pt}{FOPT} & \raisebox{-2pt}{Average}
\\[1.2pt] \hline &&&\\[-12pt]
ALEPH moments & $0.339 \,{}^{+\, 0.019}_{-\, 0.017}$ &
$0.319 \,{}^{+\, 0.017}_{-\, 0.015}$ & $0.329 \,{}^{+\, 0.020}_{-\, 0.018}$
\\[3pt]
Modified ALEPH moments  & $0.338 \,{}^{+\, 0.014}_{-\, 0.012}$ &
$0.319 \,{}^{+\, 0.013}_{-\, 0.010}$ & $0.329 \,{}^{+\, 0.016}_{-\, 0.014}$
\\[3pt]
$A^{(2,m)}$ moments  & $0.336 \,{}^{+\, 0.018}_{-\, 0.016}$ &
$0.317 \,{}^{+\, 0.015}_{-\, 0.013}$ & $0.326 \,{}^{+\, 0.018}_{-\, 0.016}$
\\[1.5pt]
$s_0$ dependence  & $0.335 \pm 0.014$ &
$0.323 \pm 0.012$ & $0.329 \pm 0.013$
\\[1.5pt]
Borel transform  & $0.328 \, {}^{+\, 0.014}_{-\, 0.013}$ &
$0.318 \, {}^{+\, 0.015}_{-\, 0.012}$ & $0.323 \, {}^{+\, 0.015}_{-\, 0.013}$
\\[2pt] \hline &&&\\[-12pt]
Average & $0.335 \pm 0.013$ & $0.320 \pm 0.012$ & $0.328 \pm 0.013$
\\[2pt] \hline
\end{tabular}
\caption{Determinations of $\alpha_{s}^{(n_f=3)}(m_{\tau})$ from $\tau$ decay data, 
in the $V+A$ channel~\protect\cite{Pich:2016bdg}.}
\label{tab:summary}
\end{table}

Ref.~\cite{Pich:2016bdg} has performed an exhaustive reanalysis of the
$\tau$ determination of the strong coupling, including many
consistency checks to assess the actual size of non-perturbative
effects. All strategies adopted in previous works and several
complementary approaches have been investigated, studying the
stability of the results and trying to uncover any potential hidden
weaknesses. Once their uncertainties are properly estimated, all
adopted methodologies result in very consistent values of
$\alpha_s(m_\tau)$. Table~\ref{tab:summary} summarizes the most
reliable and precise determinations.
The first three lines show the results obtained with different types
of (at least doubly-pinched) weights, which are sensitive to different
power corrections. The amazing stability of the fitted values reflects
the very minor numerical effect of OPE corrections, which has been
confirmed through many other additional tests. The fourth line
extracts the information from the $s_0$ dependence of a single
moment with $\omega^{(2,m)}(x) = 1 - (m+2)\, x^{m+1} + (m+1)\,x^{m+2}$, for $m=0,1,2$; although this is much more sensitive to
potential violations of quark-hadron duality, it results in values of
the strong coupling fully compatible with the
determinations in the first three lines. The fifth line adopts weights
of the form $(1-x^{m+1}) \,\mathrm{e}^{-ax}$ that suppress violations
of duality, but paying the price of a potentially larger sensitivity
to power corrections.

The overall agreement of all determinations in Table~\ref{tab:summary} shows their robustness and reliability. As expected, there is a systematic difference between CIPT and FOPT; the last column averages the results from both prescriptions, but adding in quadrature half their difference as an additional systematic error.
Averaging the five determinations, but keeping the smaller
uncertainties to account for the large correlations, one finds the
values indicated in the last line. 
From the average of the five combined (CIPT and FOPT) results in the last column, one finally gets 
\begin{equation}\label{eq:alpha-tau}
\alpha_{s}^{(n_f=3)}(m_\tau) \; =\; 0.328 \pm 0.013 \, ,
\end{equation}
in very good agreement with the ALEPH determination of the strong coupling mentioned before.

After evolution up to the scale $m_Z$, the value of $\alpha_s^{(n_f=3)}(m_\tau)$ in (\ref{eq:alpha-tau}) decreases to
\begin{equation}\label{eq:alpha-tauMZ}
\alpha_{s}^{(n_f=5)}(m_Z)\; =\; 0.1197\pm 0.0015 \, ,
\end{equation}
which nicely agrees with the direct measurement at the $Z$ peak in Eq.~(\ref{eq:alpha_Z}).
The comparison of these two determinations
provides a beautiful test of the predicted QCD running;  i.e., a very significant experimental verification of asymptotic freedom:
\begin{equation}
\left.\alpha_{s}^{(n_f=5)}(m_Z)\right|_\tau - \left.\alpha_{s}^{(n_f=5)}(m_Z)\right|_Z
\; =\; 0.0001\pm 0.0015_\tau\pm 0.0030_Z\, .
\end{equation}
Notice the order-of-magnitude difference between the errors in (\ref{eq:alpha-tau}) and (\ref{eq:alpha-tauMZ}), 
which exhibits the much higher sensitivity to the strong coupling at lower energies.

High-precision measurements of the spectral functions, especially
in the higher kinematically-allowed energy bins, would be required in
order to improve the $\alpha_s$ determination from $\tau$ decay. Both
higher statistics and a good control of experimental systematic uncertainties are
needed, which could be possible at the Belle-II experiment. An
improved understanding of higher-order perturbative corrections is
also needed.
While $\tau$ decay data are kinematically limited to $s\le m_\tau^2$,
higher values of the hadronic invariant mass can be accessed in
$e^+e^-$ annihilation. However, in the vector spectral function the onset of the QCD asymptotic behaviour is unfortunately also reached at larger values of $s$, as exhibited in Fig.~\ref{fig:SpectralFunction} for its $I=1$ component. The more
inclusive nature of the $V+A$ channel leads to a much flatter
distribution, which is related with the smaller non-perturbative
corrections to the spectral moments. Nevertheless, $e^+e^-$ data
provide useful complementary information that can be analysed through
spectral moments in the same way than with $\tau$ decay data
\cite{Eidelman:1978xy,Narison:1993sx}. The integrated distributions
provide in fact a better sensitivity to $\alpha_s$ than the ratio
$R_{e^+e^-}(s)$. However, owing to the current experimental precision,
the resulting uncertainty on $\alpha_s$ is larger than the one
achieved with $\tau$ data \cite{Boito:2018yvl}.

\section{Global PDF fits and collider measurements \hskip .2cm {\it (J. Rojo)}}
\label{sec:pdffits}

The determination of the strong coupling constant
from fits of parton distribution functions (PDFs)
has a long history.
In a joint determination of PDFs and $\alpha_s(m_Z)$, the sensitivity
to the latter arises from two different aspects of the fit.
On the one hand, through the scaling violations induced by DGLAP
evolution, which are particularly strong at small $x$.
On the other hand, from the partonic matrix
elements for those input processes that are driven by QCD
scattering already at the Born level, such as jet production.
In this context, modern PDF fits (see~\cite{Gao:2017yyd,Accardi:2016ndt,Rojo:2015acz} for recent reviews) contain
    a wide variety of hadron collider data sensitive to the value
    of the strong coupling, from inclusive jets~\cite{Rojo:2014kta}
    to direct photon~\cite{Campbell:2018wfu}, the transverse momentum of $Z$
    bosons~\cite{Boughezal:2017nla}
    and top quark pair production~\cite{Czakon:2016olj}, and therefore offer unique
    opportunities for precision determinations of $\alpha_s(m_Z)$.
    In addition, several of these processes such as inclusive jets and top quark
    production are often used as input
    for independent determinations of $\alpha_s(m_Z)$.
    Here we first review selected recent determinations of $\alpha_s(m_Z)$ in the context of global PDF fits, 
and then those determinations based on collider
    measurements that do not involve a simultaneous PDF extraction.
    
\begin{figure}[t]
  \centering
  \includegraphics[width=.42\linewidth,angle=-90]{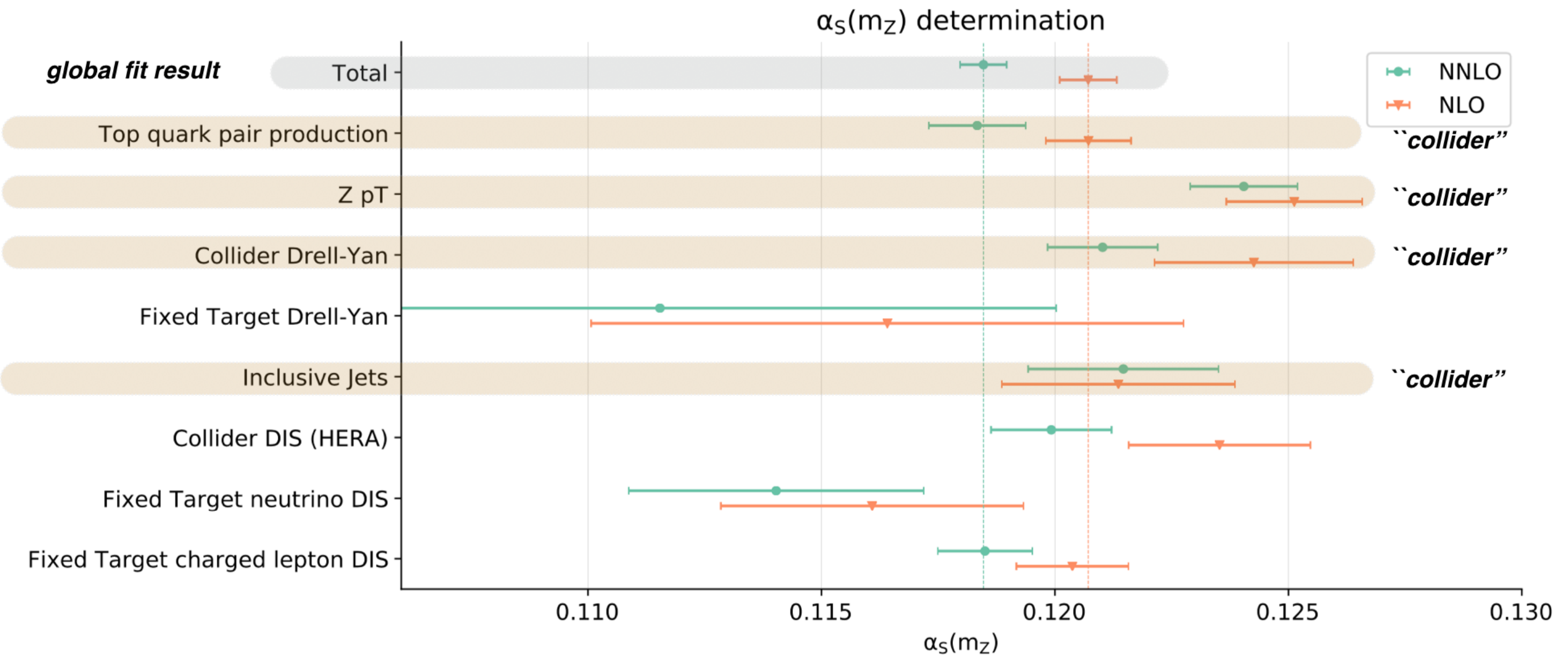}
  \caption{\small 
Modern PDF fits contain
    a wide variety of hadron collider data sensitive to the value
    of the strong coupling.
    Here we compare the values of $\alpha_s(m_Z)$ obtained
    in the NNPDF3.1 NLO and NNLO global fits with those
    obtained from the reduced $\chi^2$ restricted to specific
    families of processes.
    Several of them, such as inclusive jets and top quark
    production, are often used as input
    for independent determinations of $\alpha_s(m_Z)$.
}
\label{fig:alphas_globalfit_processes}
\end{figure}

\paragraph{Determinations within PDF fits.}
The most recent determination of $\alpha_s(m_Z)$ from a global
PDF fit is~\cite{Ball:2018iqk} based on
the NNPDF3.1 analysis~\cite{Ball:2017nwa}.
This study finds that at NNLO one has
%
\begin{equation}\label{eq:NNPDFalpha}
\alpha_s(m_Z)\,\equiv\, \alpha_s^{(n_f=5)}(m_Z)
\, =\, 0.1185\pm 0.0005_{\rm exp}\pm 0.0011_{{\rm th}}\, ,
\end{equation}
where the theory error from missing higher-order (MHO) corrections
at $\mathcal{O}\lp \alpha_s^3\rp$
is conservatively estimated from the half-shift between NLO and NNLO.
This global analysis is based on a comprehensive set of input data
from the precise HERA structure functions to jet, electroweak
gauge boson, and top quark production at the Tevatron and the
LHC.
Crucially, many of these processes exhibit a direct (and complementary)
sensitivity to $\as$.
Theoretical calculations are based on exact NNLO QCD fixed-order theory,
which for most processes used in the fit have become available
only very recently.

In Fig.~\ref{fig:alphas_globalfit_processes} we compare the values of $\alpha_s(m_Z)$ obtained
in the NNPDF3.1 NLO and NNLO global fits with those
obtained from the ``reduced'' $\chi^2$ restricted
to specific families of processes,
which we define by keeping all PDFs fixed to their best-fit values
in the global analysis, while
examining now how the $\chi^2$  contribution of a specific family varies with
the value of the strong coupling $\as$.
From this study,
one then finds that the value of $\alpha_s(m_Z)$ appears
to be determined from
    the combination of several classes of process that carry a similar
    weight in terms of their sensitivity to the strong coupling:
    $t\bar{t}$ production, $Z$ $p_T$, collider gauge boson production,
    and collider and fixed-target deep-inelastic scattering.
    In other words, there is not a single process that
    dominates the extraction of $\as$ from the global
    PDF, but rather the interplay between many of them
    with comparable weight.

\begin{figure}[t]
  \centering
  \includegraphics[width=.99\linewidth]{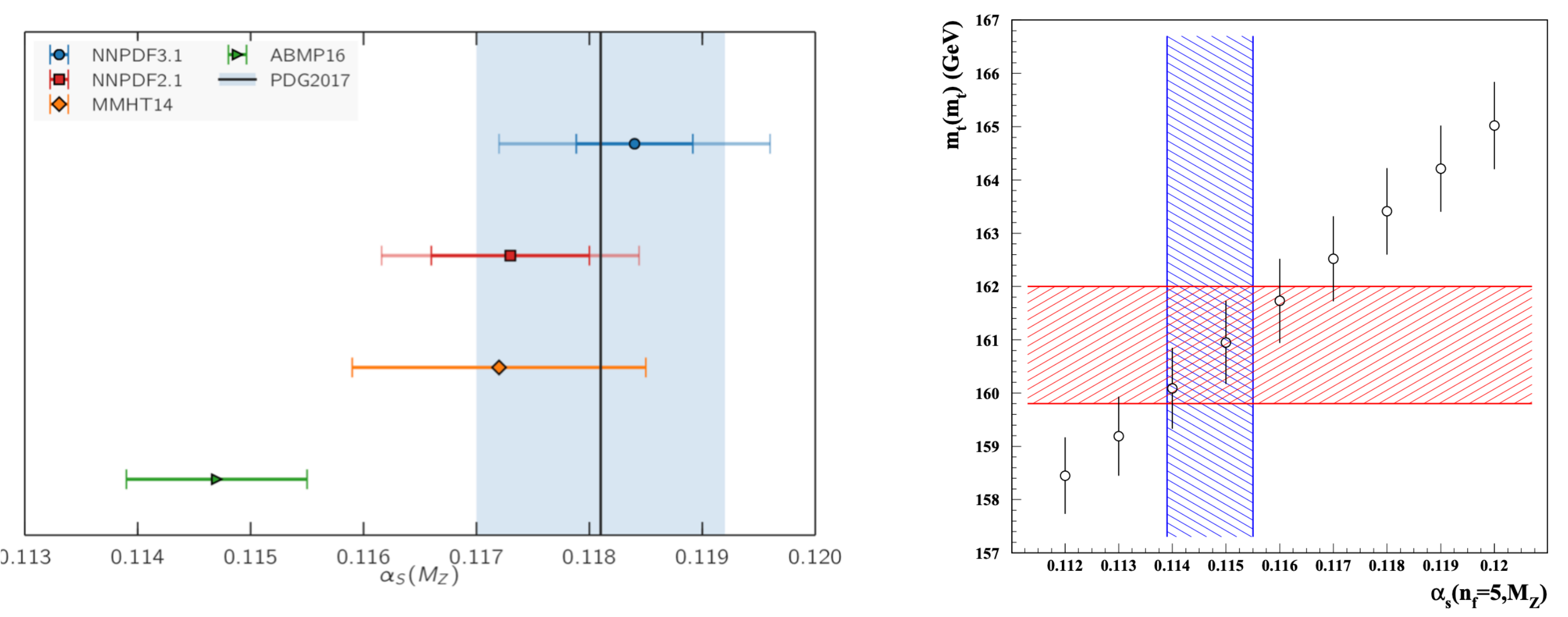}
  \caption{\small Left: comparison of the $\alpha_s(m_Z)$ values
    determined in the NNPDF3.1, NNPDF2.1, MMHT14, and ABMP16
    PDF fits compared to the PDG17 average.
    Right: the correlation between the fitted values of  $\alpha_s(m_Z)$ and $m_t(m_t)$
    obtained in the ABMP16 analysis; figure reproduced from~\cite{Alekhin:2017kpj}.
}
\label{fig:alphas_pdffit}
\end{figure}

In Fig.~\ref{fig:alphas_pdffit} (left) we compare
the $\alpha_s(m_Z)$ values
determined in the NNPDF3.1, NNPDF2.1~\cite{Lionetti:2011pw,Ball:2011us},
MMHT14~\cite{Harland-Lang:2015nxa}, and
ABMP16~\cite{Alekhin:2017kpj}
PDF fits with the current PDG average~\cite{Tanabashi:2018oca},
namely
\be
\alpha_s^{\rm (pdg)}(m_Z)=0.1181 \pm 0.0011 \, .
\ee
    For the NNPDF results, the inner error band includes
    the experimental and procedural uncertainties
    while the outer bar includes
    as well the theoretical MHO uncertainties.
    Both the NNPDF and the MMHT14 results agree within them as well
    as with the PDG averages.
    On the other hand, the ABMP16 determination is markedly lower.
    These differences can be partly explained by
    differences in the input dataset as well as in the treatment of heavy quark mass effects
    in the deep-inelastic structure functions, as discussed in~\cite{Thorne:2014toa,Alekhin:2012ig}.
    From the NNPDF3.1 result in Fig.~\ref{fig:alphas_pdffit} we can also see that, provided
    MHO uncertainties can be robustly estimated and reduced
    to a level comparable or below the experimental
    uncertainties, the determination of $\alpha_s(m_Z)$ from
    the global fit could become one of the dominant
    ingredients in the PDG average.

    In a similar way that MHO uncertainties affect the  $\alpha_s(m_Z)$ determination
    from the global fit (even though they are often neglected), one needs to
    account for the effects of other theory uncertainties such as the input
    values of the heavy quark masses.
    For example, in the description of top quark pair production cross sections,
    changes in the PDFs and $\alpha_s$ can be partially compensated by changes in
    the top quark mass $m_t$.
    This is illustrated in the right panel of Fig.~\ref{fig:alphas_pdffit}, which
    shows the correlation between the fitted values of  $\alpha_s(m_Z)$ and $m_t(m_t)$
    obtained in the ABMP16 analysis.
    As discussed in~\cite{Czakon:2016olj}, the dependence on the value of $m_t$
    can be efficiently suppressed by fitting normalised differential distributions
    where the dependence on the top quark mass cancels out.

\paragraph{Determinations from individual collider measurements.}
The same sensitivity to the value of $\alpha_s(m_Z)$ that specific collider measurements
provide within the global PDF fit (see Fig.~\ref{fig:alphas_globalfit_processes})
can also be exploited to provide independent determinations that do not
involve the simultaneous extraction of the PDFs, but that rather assume
PDFs and their uncertainties as an external input, see {\it e.g.}~\cite{Rojo:2014kta}
and references therein for the specific case of jet production.
In this context, the recent availability of the NNLO QCD corrections
for fully differential distributions in inclusive jet~\cite{Currie:2016bfm},
dijet~\cite{Currie:2017eqf}, and top quark
pair production~\cite{Czakon:2016dgf} has made possible a significant reduction of the MHO uncertainties
associated to these collider determinations, which before were limited
to NLO QCD accuracy.

\begin{figure}[t]
  \centering
  \includegraphics[width=.99\linewidth]{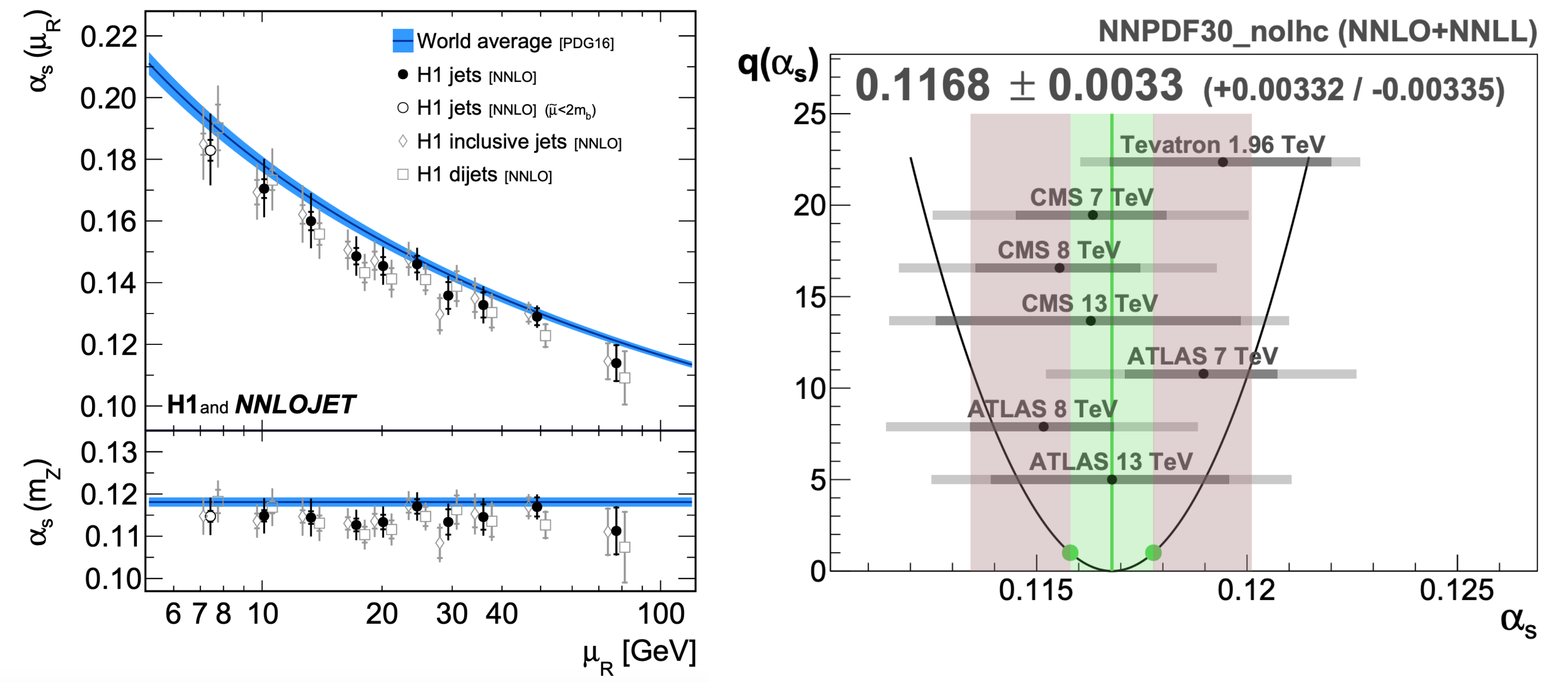}
  \caption{\small Left: summary of determinations of $\alpha_s$
    from jet production in DIS by the H1 experiment~\cite{Andreev:2017vxu} based on NNLO QCD theory.
    We show results for both $\alpha_s(\mu_R)$ and $\alpha_s(m_Z)$, where
    $\mu_R$ indicates the typical scale of the process.
    Right: results of the determination of $\alpha_s(m_Z)$ from the total $t\bar{t}$ cross section
    reported in ~\cite{Klijnsma:2017eqp} using NNPDF3.0 noLHC as input PDF set.
}
\label{fig:alphas_collider}
\end{figure}

To illustrate this point,
recently the strong coupling was determined from
jet production measurements (both inclusive jets and dijets)
in deep-inelastic lepton-proton scattering by the
H1 collaboration~\cite{Andreev:2017vxu} using the recently available NNLO
QCD calculation~\cite{Currie:2017tpe}.
Two different determinations are presented in this study: one where PDFs
and their uncertainties are taken
as external input from previous determinations
and another where $\alpha_s(m_Z)$ is fitted simultaneously with the PDFs
based on the {\tt xFitter} tool~\cite{Alekhin:2014irh}.
This analysis finds $\alpha_s(m_Z)=0.1157\pm 0.0020~({\rm exp})\pm 0.0029~({\rm th})$
and $\alpha_s(m_Z)=0.1157\pm 0.0028~{\rm (tot)}$ in each of the two cases.
Therefore despite the use of NNLO QCD calculations, theory errors from MHOs are still
significant.
In the left panel of Fig.~\ref{fig:alphas_collider} we present a
summary of the different H1 determinations of $\alpha_s$ from jet data using NNLO QCD theory.
    We show results for both $\alpha_s(\mu_R)$ and $\alpha_s(m_Z)$, where
    $\mu_R$ indicates the typical scale of the process.

    As proposed for the first time in~\cite{Chatrchyan:2013haa}, the total
    cross section for top quark pair production at the LHC also provides
    a clean observable to determine the strong coupling.
    This sensitivity has been exploited in~\cite{Klijnsma:2017eqp} to carry
    out an updated extraction of $\alpha_s(m_Z)$ from $\sigma\lp t\bar{t}\rp$
    based on data from the Tevatron Run II and the LHC 7, 8, and 13 TeV.
    In the right panel of Fig.~\ref{fig:alphas_collider} we show
    the results of this analysis using NNPDF3.0 noLHC~\cite{Ball:2014uwa} as input PDF set.
    The green line and band indicate the central value of the combination
    of all collider measurements and the associated one-sigma experimental
    uncertainty, while the red band includes as well the MHO, PDF, and $m_{t}$
    theoretical uncertainties.
    Their final combined result is $\alpha_s(m_Z)=0.1177\pm 0.0035$, in agreement
    with the PDG average.

    Other related determinations of $\alpha_s(m_Z)$ from collider
    cross sections, which do not involve simultaneously PDF extractions, are the
    CMS determination based on the $R_{3/2}$ ratio of three-to-two jet
    cross sections~\cite{Chatrchyan:2013txa} and on the
    inclusive jet cross section~\cite{Khachatryan:2014waa} at 7 TeV
    and the ATLAS determinations based on the transverse energy-energy
    correlation~\cite{Aaboud:2017fml} and the dijet azimuthal
    correlation~\cite{Aaboud:2018hie} in jet production at 8 TeV.

\begin{figure}[t]
  \centering
  \includegraphics[width=.90\linewidth]{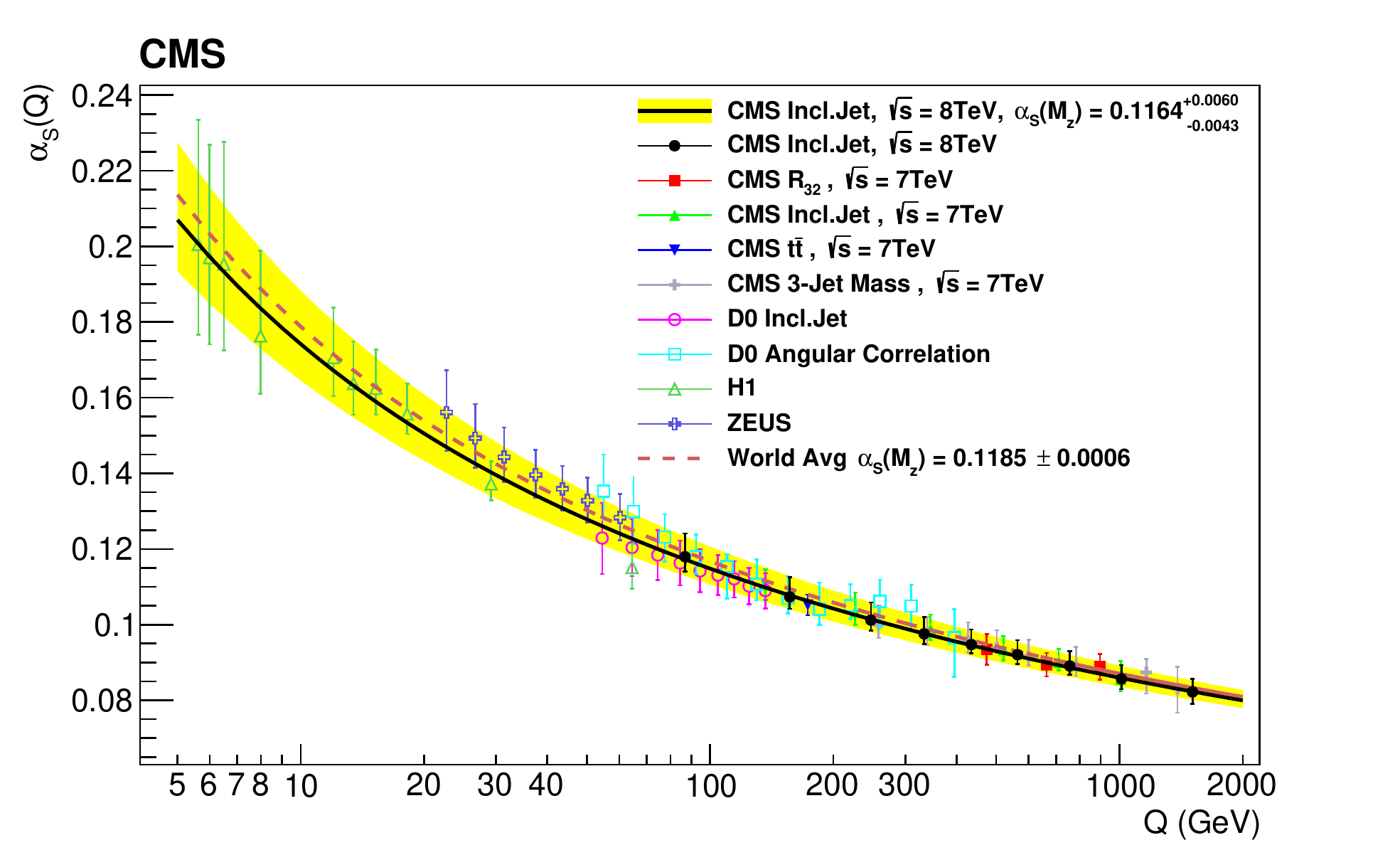}
  \caption{\small Summary of recent determinations of $\alpha_s(m_Z)$ at
    the LHC from the CMS experiment, compared to previous results
    by the Tevatron and HERA as well as with the global PDG average.
    Results are presented as a function of $Q$, the
    characteristic energy scale
    involved in each specific determination.
}
\label{fig:As_running_ALL}
\end{figure}

To provide a general overview, in Fig.~\ref{fig:As_running_ALL}
we show a summary of recent determinations of $\alpha_s(m_Z)$ at
    the LHC from the CMS experiment, compared to previous results
    by the Tevatron and HERA as well as with the global PDG average.
    Results are presented as a function of $Q$, the typical energy scale
    involved in each specific determination.
    We can see how only the LHC determinations have a direct coverage of the
    TeV region, which is also important in searches of new heavy particles
    as we discuss next.

\paragraph{Testing the running of $\bm{\alpha_s(Q)}$ in the TeV scale.}
In addition to the precision determination of the strong coupling
at the $Z$ boson mass, $\alpha_s(m_Z)$,
the direct measurement of its running with $Q$ in the TeV region
is also of high interest.
The reason is that this running is generically modified
    in the presence of new strongly interacting particles
    as predicted by many scenarios of new physics
    beyond the Standard Model (bSM), such as by squarks
    and gluinos in supersymmetry, or top partners
    in composite Higgs models.
    These modifications are for example crucial
    in scenarios where the strong and electroweak
    interactions are unified at very high scales~\cite{Dimopoulos:1981yj}.
    
\begin{figure}[t]
  \centering
  \includegraphics[width=.30\linewidth,angle=-90]{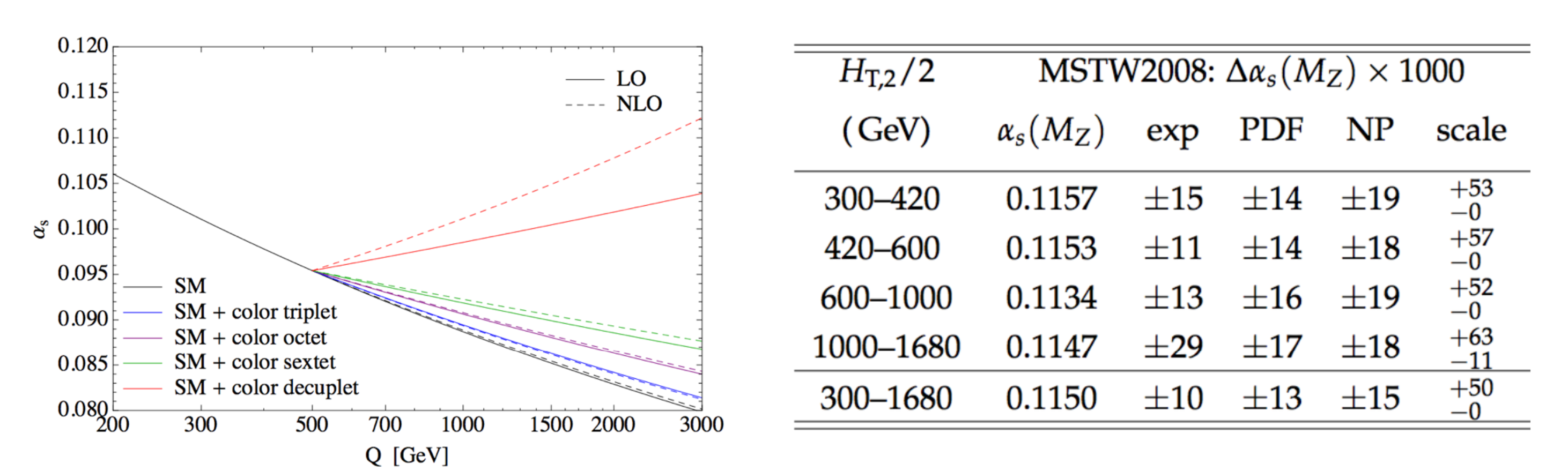}
  \caption{\small Left: the running of $\alpha_s(Q)$ is generically modified
    in the presence of new strongly interacting particles
    as predicted by many bSM scenarios, where 500~GeV mass has been assumed for the new-physics multiplets~\cite{Becciolini:2014lya}.
    Right: the running of $\alpha_s(Q)$ in the TeV scale
    can be directly determined from LHC processes
    such as multijet production, in this case
    we show the CMS 8 TeV analysis~\cite{CMS-PAS-SMP-16-008}.
}
\label{fig:alphas_running}
\end{figure}

    The effects of such new bSM
    degrees of freedom are illustrated in Fig.~\ref{fig:alphas_running},
    taken from~\cite{Becciolini:2014lya}, which
    shows how $\alpha_s(Q)$ is affected by new colored particles
    for different representations.
    In this example, a mass of 500 GeV has been assumed by the bSM particles
    belonging to new strongly interacting multiplets.
    If such new states are light enough, they could lead
    to visible effects in the running of $\alpha_s(Q)$ within the
    LHC range.
    Conversely, precision measurements of $\alpha_s(Q)$ in the
    TeV range could be of use to derive stringent model-independent
    bounds on bSM scenarios containing strongly interacting sectors.
    
Direct measurements of $\alpha_s(Q)$ in the TeV range can be obtained
from processes such as inclusive jet, dijet, and multijet production,
as well as top quark pair production in the tail of
the $m_{t\bar{t}}$ distribution,
which are both sensitive to the
value of the strong coupling and lead to sizable event rates
in the TeV region.
For instance, in Fig.~\ref{fig:alphas_running} (right) we show
the CMS measurement of $\alpha_s(Q)$ from the multijet
cross sections at 8 TeV~\cite{CMS-PAS-SMP-16-008} using
NLO QCD calculations and MSTW20018 as input PDF set.
By restricting the input data used in the fit
to separate bins of the $H_{T,2}/2$ kinematic variable,
one can effectively measure $\alpha_s(Q)$ for increasing values of
$Q$ and thus validate its running by comparing it
to the SM predictions.
While the current results are limited by the scale uncertainties of the NLO
calculation as well as by statistics in the TeV region, future extractions based
on NNLO QCD and in a much higher integrated luminosity should be in the position
of providing stringent tests of bSM scenarios where the running of $\alpha_s(Q)$
is modified as compared to the QCD prediction.

\paragraph{$\bm{\alpha_s(m_Z)}$ from electron-positron collisions.}
The strong coupling can also be determined from high-energy $e^+e^-$ collisions, by
exploiting how the pattern of QCD radiation is modified in these collisions once $\alpha_s(m_Z)$
is varied.
Specifically, a number of the so-called event shapes such as the thrust or the $C$-parameter
have been in the last years used to extract $\alpha_s(m_Z)$ from LEP measurements.
The main benefit of this process is that it is experimentally very clean, and
that one has good control over the perturbative calculation including resummation,
with recent extractions being based on NNLO+N$^3$LL theory.
On the other hand, the extraction of $\alpha_s(m_Z)$ from event shapes requires a precise modeling
of the hadronisation mechanism of quarks and gluons (and related non-perturbative effects) in order
to connect the perturbative calculation with the LEP cross sections.

Two recent determinations of the strong coupling
constant from event shapes~\cite{Hoang:2015hka,Abbate:2010xh},
based on the Soft-Collinear Effective Theory (SCET) formalism,
find the following results:
\be
\alpha_s(m_Z) = 0.1135\pm 0.0011  \qquad {\rm (Thrust)} \, ,
\label{eq:thrust}
\ee
\be
\alpha_s(m_Z) = 0.1123\pm 0.0015  \qquad {\rm (C-parameter)} \, ,
\label{eq:C-par}
\ee
so rather smaller than the PDG average.
The reason for these differences is still under investigation.
One possible reason could be related to the treatment of non-perturbative effects,
which are more challenging to control than the perturbative part
of the calculation where residual MHO corrections appear to be 
quite small.

\section{QCD static energy \hskip .2cm {\it (A. Vairo)}}
\label{sec:static}

Lattice QCD provides a potentially very accurate source of $\alpha_s$~\cite{Aoki:2016frl}, see \sect{sec:lattice}.
Indeed among the determinations quoted by the Particle Data Group~\cite{Tanabashi:2018oca}, the most precise ones are lattice determinations.
Among the lattice determinations, the extraction of $\alpha_s$ from the QCD static energy 
is particularly attractive as the perturbative expansion of this quantity is very accurately known as well as its lattice determination.  

Lattice determinations of the QCD static energy have started since the inception of lattice QCD itself. 
The QCD static energy, $E_0(r)$, is the energy possessed by a static quark and a static antiquark located at a distance $r$. 
Its expression in Minkowski spacetime is~\cite{Wilson:1974sk,Susskind:1976pi,Fischler:1977yf,Brown:1979ya}
\begin{equation}
E_0(r) = \lim_{ T\to\infty}\frac{i}{T} \ln \, \left\langle {\rm Tr} \,{\rm P} \exp\left\{i g \oint_{r\times T} dz^\mu \, A_\mu(z)\right\} \right\rangle, 
\label{E0}
\end{equation}
where the integral is over a rectangle of spatial length $r$ and time length $T$; 
$\langle \dots \rangle$ stands for the path integral over the gauge fields $A_\mu$ and the light-quark fields, 
P is the path-ordering operator of the color matrices (fields are time ordered) eventually traced and $g$ is the QCD gauge coupling ($\alpha_s = g^2/(4\pi)$). 

In the short range $r\Lambda_{\rm QCD} \ll 1$, for which $\alpha_s(1/r) \ll 1$, $E_0(r)$ may be computed as a perturbative expansion in $\alpha_s$ 
(evaluated at a typical scale of order $1/r$; at three loops and higher, however, also couplings at the lower energy scale $\alpha_s/r$ show up):
\begin{equation}
E_0(r) = \Lambda - \frac{4\alpha_s}{3r}(1 + \dots),
\label{E0PT}
\end{equation}
where $\Lambda$ is a constant 
and the dots stand for higher-order terms.
The expansion of $E_0(r)$ in powers of $\alpha_s$ has been computed at order $\alpha_s^4/r \times \ln \alpha_s$ in~\cite{Brambilla:1999qa,Brambilla:1999xf},
at order $\alpha_s^4/r$ in~\cite{Anzai:2009tm,Smirnov:2009fh},
at order $\alpha_s^5/r \times \ln \alpha_s$ in~\cite{Brambilla:2006wp},
all orders $\alpha_s^{4+n}/r \times \ln^{1+n} \alpha_s$ (N$^2$LL accuracy) have been computed in~\cite{Pineda:2000gza},
and all orders $\alpha_s^{5+n}/r \times \ln^{1+n} \alpha_s$ (N$^3$LL accuracy) have been computed in~\cite{Brambilla:2009bi}.
A compact summary of the perturbative expression of $E_0(r)$ can be found in~\cite{Tormo:2013tha}.
 
In lattice regularization $\Lambda$ includes a linear divergence due to the 
self-energy $\sim\as(1/a)/a$, with $a$ the lattice spacing.
In dimensional regularization the linear divergence vanishes, 
but the perturbative expansion of $E_0(r)$ is affected by a renormalon ambiguity of order $\Lambda_{\rm QCD}$~\cite{Pineda:1998id,Hoang:1998nz}.
The renormalon ambiguity reflects in the poor behaviour of the perturbative series, which 
may be cured by subtracting the renormalon from the perturbative series and suitably redefining the constant $\Lambda$.

If we are interested in \eqref{E0PT} just for extracting $\alpha_s$,
then the relevant information is encoded in the slope of the static energy,  i.e., the force 
\begin{equation}
F(r) = \frac{d}{d r}E_0(r).
\label{force}
\end{equation}
The force does not depend on $\Lambda$. 
In lattice regularization it is not affected by the linear self-energy divergence and 
it is free from the renormalon of order $\Lambda_{\rm QCD}$.
The continuum perturbative expansion of the force shows, therefore, a convergent behaviour (at least up to three loops).
In order to compute $\alpha_s$, one may compare the perturbative continuum expression of the force with its lattice determination.
However, a precise direct lattice computation of the force is challenging (see also below),
while an accurate determination of $E_0(r)$ is much easier, for it amounts at extracting the exponential fall off of a static Wilson loop.
Hence an alternative strategy is to integrate the perturbative continuum expression of the force over the distance,
obtaining back the static energy (see also~\cite{Necco:2001gh}),
\begin{equation}
E_0(r)=\int_{r_*}^{r}dr'\, F(r'),
\label{E0force}
\end{equation}
up to an irrelevant constant determined by the arbitrary distance $r_*$, 
which can be reabsorbed in the overall normalization when comparing finally with lattice data.
Equation \eqref{E0force} amounts effectively to a rearrangement of the perturbative expansion of the static energy.
The integration in \eqref{E0force} can be done (numerically) keeping the strong coupling
that appears in the perturbative expansion of the force running at a natural scale of the order of the inverse of the distance. 

Whatever strategies one pursues, the comparison of the perturbative expression of the QCD static energy or the force
to lattice QCD determinations provides with (at least) three-loop accuracy the QCD strong coupling $\alpha_s$
at a typical scale that is large enough for perturbative QCD to work and smaller than the inverse lattice spacing $1/a$.
This means that the distance range of the QCD static energy explored by lattice QCD that is relevant for the extraction
of $\alpha_s$ goes from about 0.2~fm to about 0.05~fm or slightly below.
Hence, the QCD static energy provides an accurate determination of $\alpha_s$ at a low energy scale,
typically about 1.5~GeV, for which it has very few competitors (one of these is the extraction of $\alpha_s$ from the $\tau$ decay, 
as the mass of the $\tau$ is about $1.777$~GeV,
see Sect.~\ref{subsec:Tauwidth}).
This gives to this determination a value in itself and adds to its relevance, 
regardless of the final precision in the determination of $\alpha_s(m_Z)$.
Nevertheless, the precision in the final determination of $\alpha_s(m_Z)$ is also competitive with other determinations.
One of the challenges in extracting $\alpha_s$ from the QCD static energy is to move the distance range towards shorter ones 
in order to be deeper in the perturbative regime and reduce uncertainties originating from not included higher-order terms,
keeping at the same time sufficiently  away from the minimal lattice distance, $a$.

\paragraph{Status.}
Precision determinations of $\alpha_s$ from the QCD static energy have started in quenched QCD with Ref.~\cite{Necco:2001gh}.
The precision of the data allowed computing numerically the derivative of the static energy and hence the force: $\alpha_s$ was computed from the force. 
More recent pure gauge theory ($\Nf=0$) determinations can be found in~\cite{Brambilla:2010pp,Husung:2017qjz}.

Static-energy determinations with 2 flavors are in~\cite{Jansen:2011vv,Karbstein:2014bsa,Karbstein:2018mzo} 
and with 2+1 flavors in~\cite{Bazavov:2012ka,Bazavov:2014soa}.
A summary of these latest results is in~\cite{Vairo:2016pxb}.
The most recent analysis of the static energy with 2+1 flavors is in~\cite{Takaura:2018lpw,Takaura:2018vcy}. 
Considering (somewhat long) distances up to about 0.35~fm to extract $\alpha_s$, 
it obtains results that are consistent with the ones in~\cite{Bazavov:2012ka,Bazavov:2014soa}, but with larger final errors, see 
also \fig{alphasMSbarZ}.

\begin{figure}[ht]
\makebox[0cm]{\phantom b}\put(0,0){\epsfxsize=7truecm \epsffile{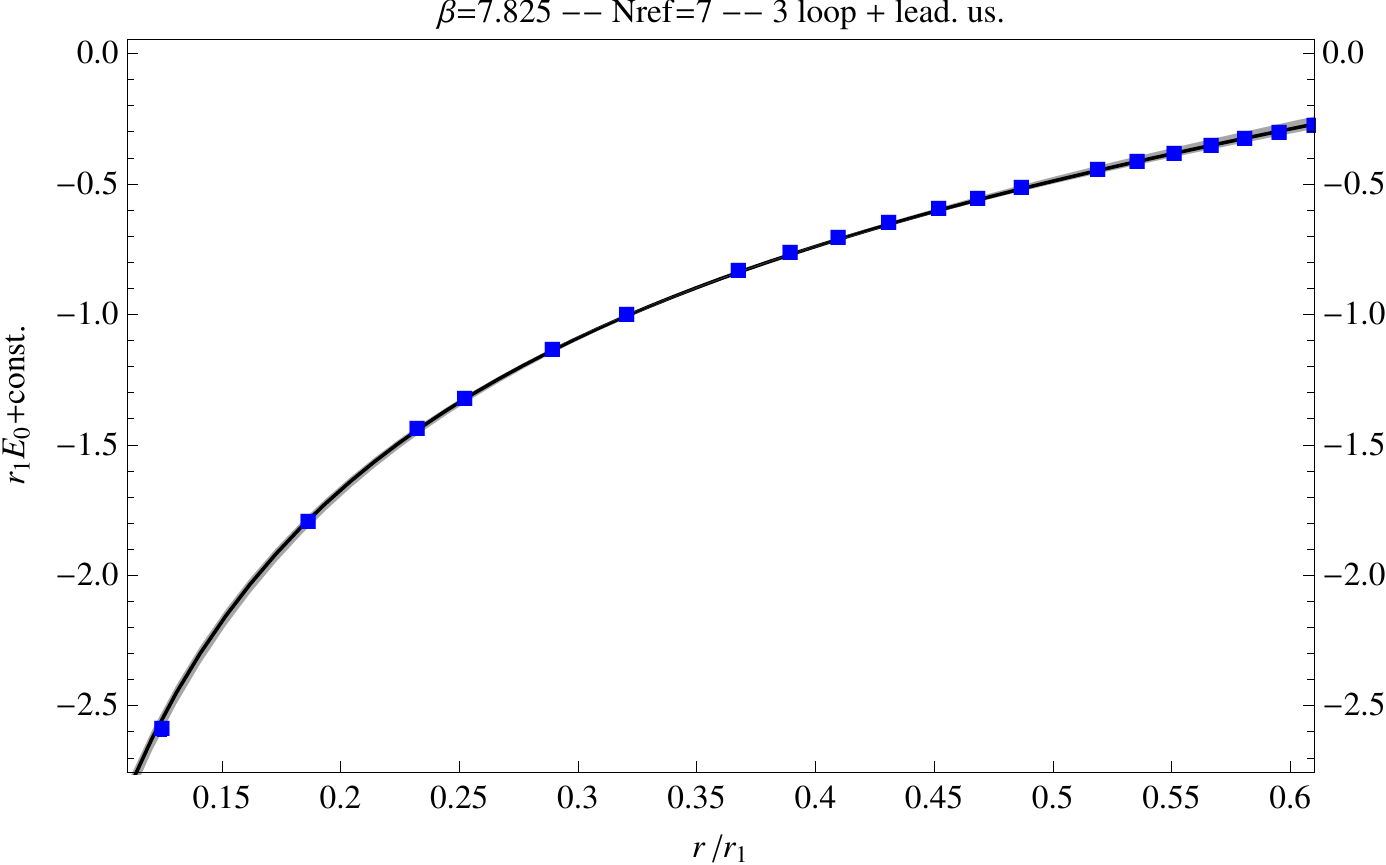}}\put(220,0){\epsfxsize=7truecm \epsffile{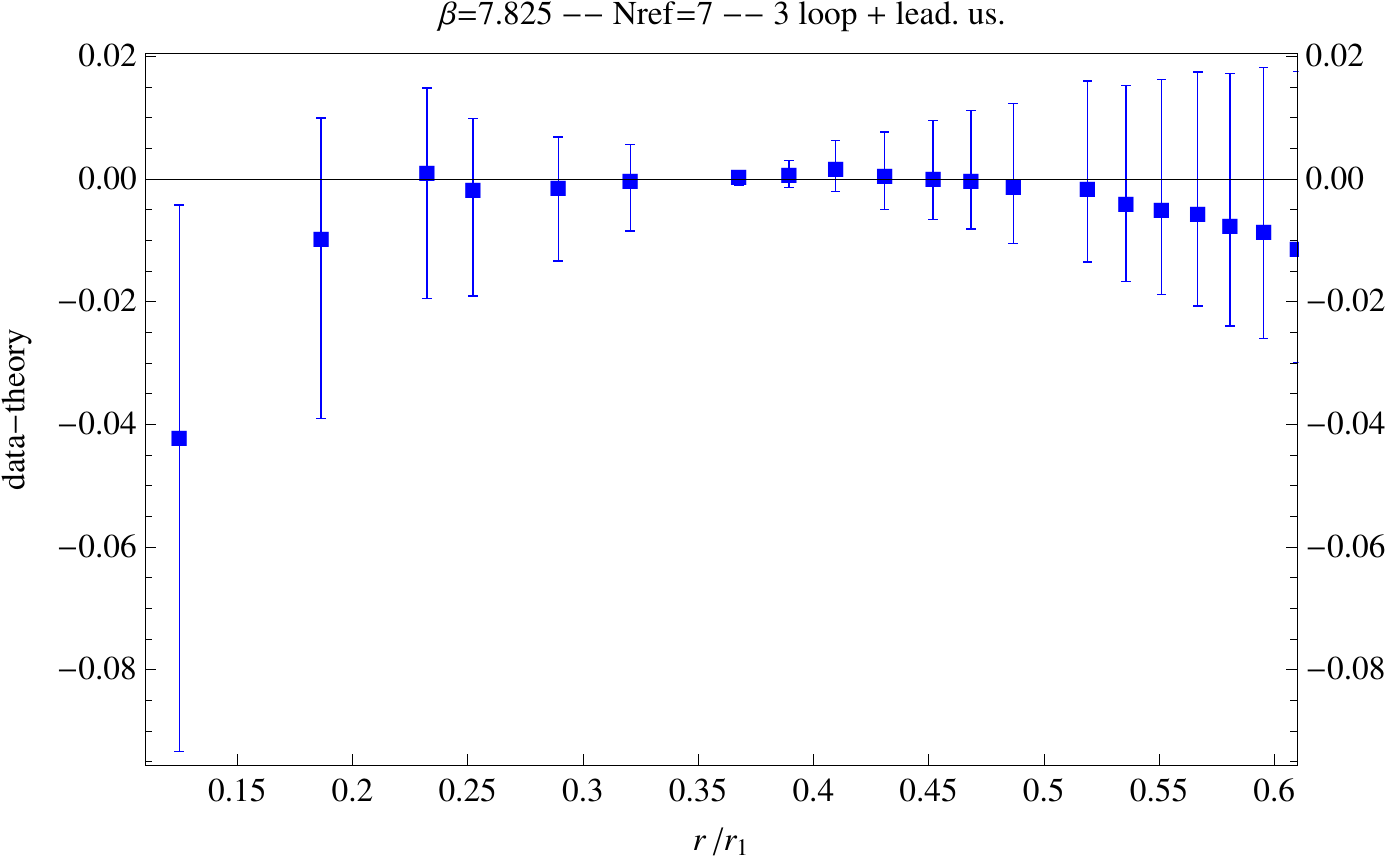}}
\caption{Left panel: Lattice data of the QCD static energy for $\beta=7.825$ compared with the
  N$^2$LL perturbative expansion (equivalent to 3 loops plus leading ultrasoft resummation) evaluated for $r_1\Lambda_{{\overline{\rm MS}}}=0.495^{+0.028}_{-0.018}$
 (the thin grey band reflects the uncertainty). Right panel: The same lattice data with the perturbative expression subtracted.  
  The error bars are obtained by adding, in quadrature, the errors of the lattice data and the uncertainty of the perturbative
  expression due to the variation of $r_1\Lambda_{{\overline{\rm MS}}}$.
  The normalization constant in the difference between lattice and perturbative data has been fixed on the seventh point.
  Data from~\cite{Bazavov:2014soa}.
  \label{fig1}}
\end{figure}

The determination of~\cite{Bazavov:2014soa} is based on the perturbative expression~\eqref{E0force} computed up to N$^2$LL accuracy and
compared with 2+1 flavors lattice data (strange quark mass at its physical value and pion mass at 160~MeV) 
for $\beta = 7.373$, $7.596$ and $7.825$ in the range from 0.05~fm to 0.14~fm. 
The result is 
\begin{equation}
  r_1\Lambda_{{\overline{\rm MS}}}=0.495^{+0.028}_{-0.018}\,.
\end{equation}
Taking the lattice scale $r_1 = 0.3106 \pm 0.0017$~fm from~\cite{Bazavov:2010hj}, the result translates into 
\begin{equation}
  \Lambda_{{\overline{\rm MS}}}  = 315^{+18}_{-12}~{\rm MeV}\,
\end{equation}
for three flavors. The above value of $\Lambda_{{\overline{\rm MS}}}$ converts into  
\begin{equation}\label{eq:StaticEalpha}
\alpha_s(1.5\,\textrm{GeV},n_f=3)=0.336^{+0.012}_{-0.008}\,,
\end{equation}
at the energy scale of $1.5~$GeV that corresponds to a distance of about 0.13~fm and is about the largest scale consistent with a three-flavors running.
Finally, this value of $\alpha_s$ evolves to the value 
\begin{equation}
  \alpha_s  (m_Z,n_f=5)=0.1166^{+0.0012}_{-0.0008}\,,
\label{alphasMZ}
\end{equation}
at the $Z$ mass. A comparison of the lattice data for $\beta=7.825$ with the perturbative expression is shown in Fig.~\ref{fig1}.
In the considered distance range, non-perturbative corrections, which would manifest themself as power corrections in $r$,
are consistent with zero.

A similar analysis has been done in~\cite{Bazavov:2014soa} with the force.
However, the numerical derivatives of the static energy add to the uncertainties of the strong coupling determination,
which, done in this way, turns out to be consistent, but less accurate, than the direct determination from the static energy presented above.

\paragraph{Perspectives.}
The determination of $\alpha_s$ from the QCD static energy has still room for improvement
and will face in the near future interesting challenges. In the following we list some of them.

Not all of the available theoretical information on the short-distance static energy has been used:
in particular, presently available data do not seem sensitive neither to the N$^3$LL expression of the static energy, nor to short-range non-perturbative effects
(e.g., condensates $\sim r^3 \langle g^2{\bf E}(0)^2\rangle$, or correlators $\displaystyle \sim r^2 \int_0^\infty dt \langle g{\bf E}(t) \cdot g{\bf E}(0)\rangle$, 
where ${\bf E}$ is the chromoelectric field).
One question is then if lattice data at shorter distances and more accurate will become sensitive to these effects.
It should be mentioned that an assessment of the size of the non-perturbative contributions to the static energy will have a major impact on quarkonium physics.

Forthcoming lattice computations with 2+1+1 flavors will naturally raise the question of how much the data for the static energy will turn out to be sensitive to the charm mass
and of how much this will affect the extraction of $\alpha_s$.
The region around the charm quark mass, $(1.5\,\text{GeV})^{-1} \approx 0.13$\,fm, and below is indeed the region from where $\alpha_s$ from the static energy mostly comes.

More precise lattice data may also allow for a competitive determination of $\alpha_s$ directly from the force,
as it was done long ago in quenched QCD~\cite{Necco:2001xg}, and as it may be possibly done also by looking to loop functions different from the static Wilson loop.

\paragraph{$\boldsymbol{\alpha_s}$ from the static energy at very short distances.}
As mentioned above, one may look at the static energy at very short distances to minimize the effect of unknown higher-order terms,
which include perturbative and non-perturbative contributions, and to have a better  behaved series.
This may certainly represent an improvement with respect to present determinations, but one should also realize that 
the range from 0.05~fm to 0.14~fm used in~\cite{Bazavov:2014soa} already shows a well behaved perturbative series, 
agreement with data and a stable result with respect to (adding/subtracting) lattice data, renormalization scale, etc.

A new set of data from a short-distance determination of the static energy in the range from about 0.02~fm to 0.18~fm 
on a lattice with 2+1 flavors will be presented in Ref.~\cite{Bazavov:2018}.
At short distances finite volume effects are irrelevant, as confirmed by previous analyses.
A possible issue is however that at small lattice spacings the Monte Carlo evolution of the topological charge freezes.
Although a thoroughly investigation is still to be done, the analysis performed in~\cite{Bazavov:2017dsy} does not show
for the considered observables any sensitivity if the topological charges 0, 1 and 2 are considered.

\begin{figure}[ht]
\makebox[0truecm]{\phantom b}
\put(100,0){\epsfxsize=8.5truecm \epsffile{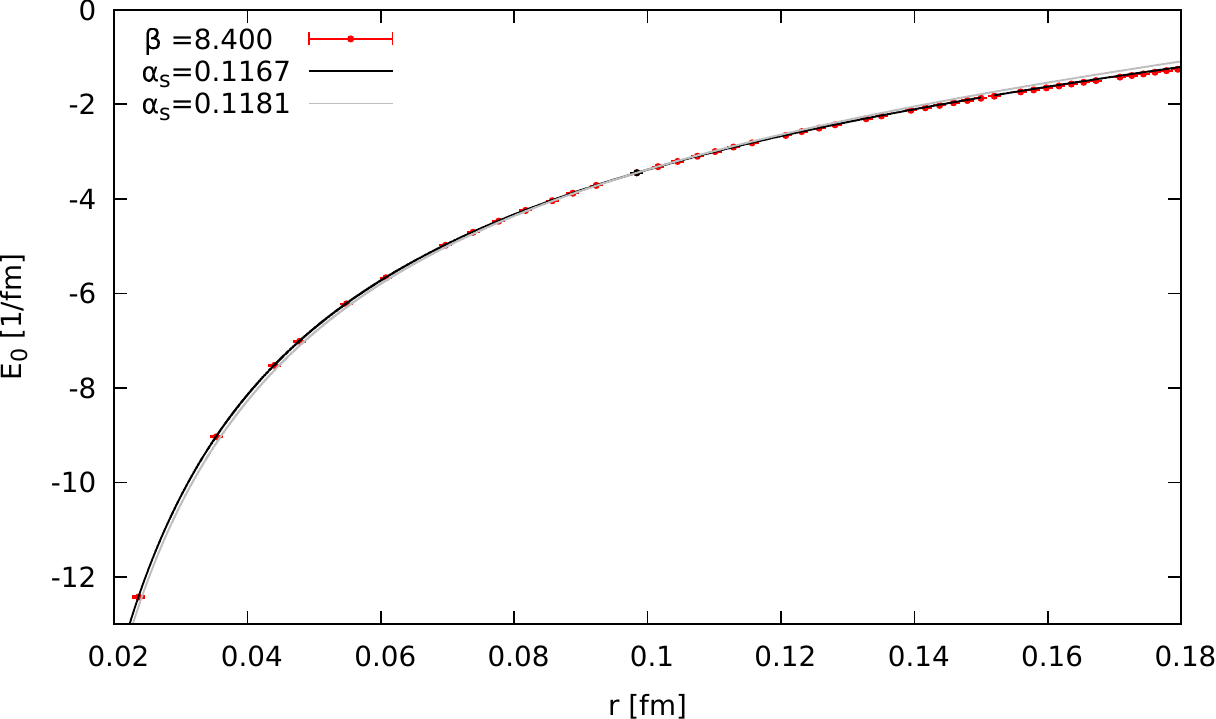}}
\caption{Preliminary lattice data for the static energy at $\beta = 8.4$ from~\cite{Bazavov:2018} compared with the N$^2$LL perturbative expression, 
  for $\alpha_s(m_Z,n_f=5)=0.1167$ and for $\alpha_s(m_Z,n_f=5)=0.1181$ (PDG value~\cite{Tanabashi:2018oca}).
\label{fig2}}  
\end{figure}

Preliminary results are shown in Fig.~\ref{fig2}. 
The data prefer a somewhat lower value of $\alpha_s$ with respect to the PDG average, but consistent with the result \eqref{alphasMZ}.

\paragraph{Charm mass effects.}
Since the charm mass is larger than $\Lambda_{\rm QCD}$, at distances comparable with the inverse of the charm mass or shorter, charm-mass effects may be computed in perturbative QCD.
Indeed, the effect of a charm mass loop to the QCD static energy is well known (see Ref.~\cite{Eiras:2000rh}).
As mentioned above, a finite charm mass contributes to the static energy at distances where one mostly compares lattice data with perturbative QCD, and eventually extracts $\alpha_s$. 
Similarly, charm mass loops are relevant for precision bottomonium physics, because the typical momentum transfer inside the $\Upsilon(1S)$ is about the charm mass.

\begin{figure}[ht]
\makebox[0truecm]{\phantom b}
\put(80,0){\epsfxsize=8.0truecm\epsffile{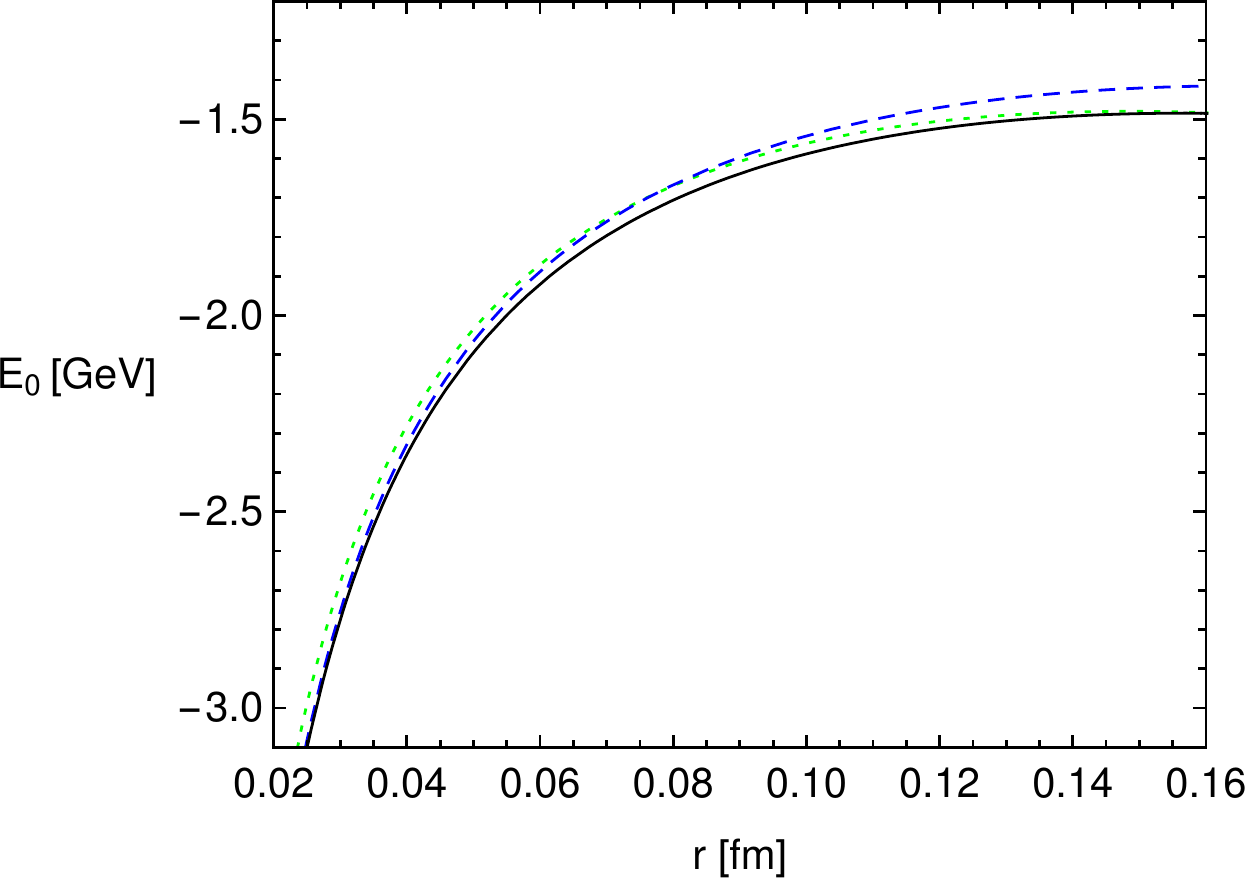}}
\caption{Dashed blue: massless charm (case of 4 active flavors); dotted green: infinitely heavy charm (case of 3 active flavors);
black: massive charm at 1 loop ($\overline{\rm MS}$ mass taken to be $1.237$~GeV). 
\label{fig3}}
\end{figure}

In Fig.~\ref{fig3}, we compare at one loop the static energy in perturbative QCD with $\alpha_s$ running with 3 flavors, 4 flavors 
and with 4 flavors plus the one-loop contribution of a massive charm. At large distances, the charm decouples, 
and the static energy is effectively  described by a 3 flavors $\alpha_s$. 
At short distances, the charm may be considered massless, and the static energy is effectively  described by a 4 flavors~$\alpha_s$. 
Accurate lattice data should be sensitive to this transition and, in particular, to the intermediate region where the static energy is best described by a charm with a finite mass.

\paragraph{$\boldsymbol{\alpha_s}$ from the force.}
Precise lattice data of the QCD static energy may allow for a precise determination of the force by interpolating the lattice points and 
numerically performing the derivatives. An alternative direct determination of the force could come from computing 
a Wilson loop with a chromoelectric field insertion~\cite{Brambilla:2000gk,Vairo:2016pxb}:
\begin{equation}
F(r) = - \lim_{ T\to\infty}\frac{ \left\langle {\rm Tr} \, {\rm P} \,\hat{\bf r} \cdot g{\bf E}(t,{\bf r}) 
\exp\left\{\displaystyle i g \oint_{r\times T} dz^\mu \, A_\mu(z)\right\} \right\rangle}
{\left\langle {\rm Tr} \, {\rm P} \exp\left\{\displaystyle i g \oint_{r\times T} dz^\mu \, A_\mu(z)\right\} \right\rangle}\,.
\end{equation}
This quantity has been computed in the context of determining the quarkonium potentials on the lattice in Ref.~\cite{Koma:2006si}. 
However, a dedicated study with the aim of extracting $\alpha_s$ is still to be done. 
In particular, the lattice version of $g{\bf E}(t,{\bf r})$ needs to be properly renormalized.

\section{Lattice determinations \hskip .2cm {\it (R. Sommer)}}
\label{sec:lattice}

Lattice gauge theory is a non-perturbative formulation of 
QCD, which allows to evaluate the Euclidean path integral
by a Monte Carlo ``simulation''.
The process starts from discretizing space-time
on a hypercubic lattice with lattice spacing $a$
and thus a momentum cutoff of $|p_\mu|\leq \pi/a$. 
Due to the large number of degrees of freedom and 
the fermionic nature of quark fields, the 
Monte Carlo process requires considerable effort
and both the simulated space-time $T\times L^3$ and
the spacing $a$ are restricted. 
Although lattice ensembles with $L/a =\rmO(100)$ are possible nowadays, even this number sets
limits to what can be done, which we will discuss shortly. 

First, let us give a broad characterization of
the lattice methods for extracting the QCD coupling.
Indeed there are several methods. Their differences are at least as
big as the difference of a phenomenological extraction from DIS vs. from 
$\tau$ decays. It is important and not too difficult to have a 
rough grasp of these differences. For more details we refer
to~\cite{Aoki:2016frl}.

In general, one performs a Monte Carlo evaluation of the path integral
representation of observables (correlation functions) for 
a few suitably chosen values of 
\begin{equation}
 L/a,\; T/a,\;g_0,\;\{ a m_i, \;i=1\ldots \Nf\} \,.
\end{equation}
Here $L/a$ is the number of points of the world in each space dimension,
$T$ (often bigger than $L$) is the extent of the time axis,  
$g_0$ is the bare coupling of the theory and $a m_i$ are the bare 
quark masses. Think now of the theory with just 
the lightest three quarks and isospin symmetry, $m_1=m_2$. The first step is to 
obtain the relation between the bare 
parameters, $g_0,\,am_1,\,am_3$ and three
hadronic, low energy quantities, which are conveniently chosen to be
the masses of pion and kaon as well as a 
leptonic decay constant, which we here just take to be $f_\pi$.\footnote{Pion and kaon masses are clearly the 
observables of choice to fix the light quark masses. However, for  
fixing the bare coupling or equivalently the overall scale of QCD,
the nucleon mass would be more natural. Unfortunately it is difficult 
once precision is required. See \cite{Sommer:2014mea} for details.}
More precisely, the functions
\bea
     F_1(g_0,am_1,am_2) &=& \frac{m_\pi}{f_\pi}\,,
     \\
     F_2(g_0,am_1,am_2) &=& \frac{m_\mathrm{K}}{f_\pi}\,,
     \\
     F_3(g_0,am_1,am_2) &=& a\,f_\pi\,,     
\eea
are determined for a few values of their arguments.
Then, at fixed $g_0$, one finds by appropriate fits/extrapolations values 
$\mu_i^*(g_0)$ such that
$F_1(g_0,\mu_1(g_0),\mu_2(g_0)) = {m_\pi^\mathrm{QCD}} / {f_\pi^\mathrm{QCD}}$ and analogous for $F_2$. 
Here the label QCD refers to PDG numbers corrected for 
isospin violating and electromagnetic effects. 
$F_3$ is then used to determine the lattice spacing, 
$a(g_0) = F_3(g_0,\mu_1(g_0),\mu_2(g_0)) / f_\pi^\mathrm{QCD}$. 
The latter step is called scale setting because from now on all dimensionful quantities which are originally just defined in units of the lattice spacing $a$ can now be expressed in
physical units. All of this has to be done in volumes
which are large enough such that finite size effects on the used
hadron masses and decay constants are negligible. Fortunately finite size effects
are quite well understood and asymptotically decrease exponentially 
$\sim \exp(-m_\pi L)$. Choosing $m_\pi L \gsim 4$ is sufficient. 
Such a bound imposes a limitation to the lattice spacings available in large
volume. For numbers one has to take into account that simulations can also
be carried out at pion masses larger than in Nature. What is roughly realized nowadays is
\be
    a \gsim 0.04 \fm  \label{eq:largev}
\ee
in a ``large volume''.

\subsection{Methods for the strong coupling.}

The general method for extracting $\alpha_\msbar$
with the help of lattice QCD is to consider 
a short-distance, one-scale, observable with an
expansion (an $\alpha_{\overline{\rm MS}}(\mu)^0$ term
can always be subtracted)
\begin{eqnarray}
   {\oO}(\mu) = c_1 \alpha_{\overline{\rm MS}}(\mu)
              +  c_2 \alpha_{\overline{\rm MS}}(\mu)^2 + \cdots \,.
\label{eq:alpha_series}
\end{eqnarray}
The observable is computed by lattice QCD and
applying the perturbative expansion with $c_{n_\mathrm{l}+1}$ 
known is denoted by a determination at 
$n_\mathrm{l}$ loops.  It means that for the effective coupling 
\begin{eqnarray}
   \alpha_{\rm eff} =  {\oO}/c_1 \,
   \label{eq:alpeff}
\end{eqnarray}
the $n_\mathrm{l}+1$ loop $\beta$-function is known. 
We will mostly have $n_\mathrm{l}=2$, but for the static potential $n_\mathrm{l}=3$ (with some complications, see Sect.~\ref{sec:static}) holds.

\paragraph{Advantages.}
An important advantage of taking $\oO$ from lattice QCD compared to
using experimental data is that one is automatically in the Euclidean
where perturbation theory works and no hadronisation corrections, duality violations etc. are a concern. 
Furthermore one has the freedom to design suitable observables.

\paragraph{Disadvantages.}
In practice, lattice QCD simulations are restricted to $\Nf=4$ quarks at the most,
because the b-quark is simply too heavy. One relies 
on including the heavy flavors, mostly including the charm, by perturbation theory, using 
4-loop matching and 5-loop running in the $\msbar$ 
scheme~\cite{Bernreuther:1981sg,Chetyrkin:2005ia,Schroder:2005hy,Kniehl:2006bg,vanRitbergen:1997va,Czakon:2004bu,Luthe:2016ima,Herzog:2017ohr,Baikov:2016tgj,Grozin:2011nk}. Worries that this might compromise the error estimates of the results have recently been removed by a non-perturbative study~\cite{Athenodorou:2018wpk}. 

We turn to the announced different categories.

\paragraph{(I) Continuum-limit observables in large volume.}
\begin{figure}[t]
  \centering
  \includegraphics[width=.95\linewidth,angle=0]{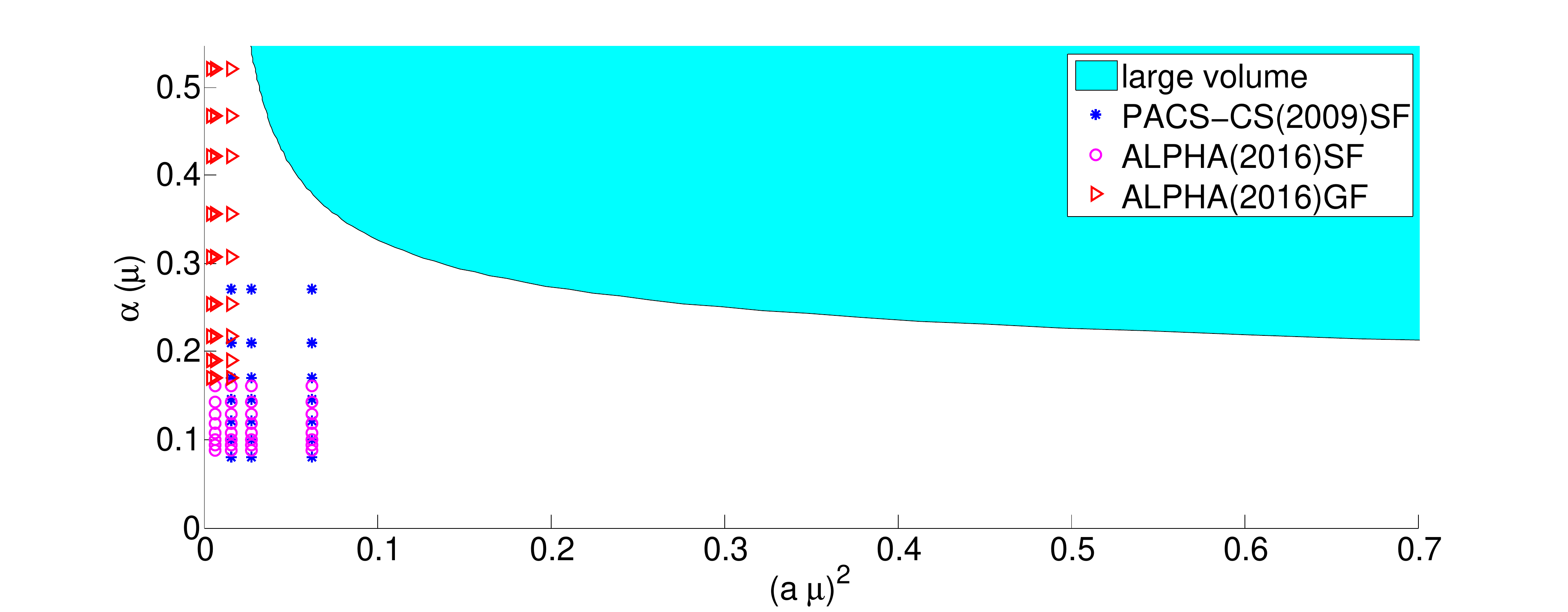}
  \caption{\small The plane $\alpha_\msbar^{(3)}(\mu)$
  against the scale $\mu$ in lattice units. $a$ is the lattice spacing and 
  the blue region corresponds to the rough bound $a>0.04\,\fm$. Note that the
  continuum limit is approached by extrapolations with 
  $a\mu\ll 1$. The points on the left correspond to actual Monte Carlo simulations in category (III).
}
\label{fig:alpha_lattice_landscape}
\end{figure}

The most straightforward strategy is to take 
a finite observable $\oO$ which depends on a single 
large momentum scale $\mu$ and which is defined in a 
large volume (i.e., large enough such that finite volume effects can be neglected). One can then (and needs to) take the continuum limit 
\begin{eqnarray}
   {\oO}(\mu) \equiv \lim_{a\rightarrow 0} {\oO}_{\rm lat}(a,\mu) 
              \mbox{  with $\mu$ fixed}\,.
\end{eqnarray}
In practice this is done by an extrapolation in 
$a\mu \to 0$ using the structural information from
Symanzik's effective theory \cite{Symanzik:1980UH} of lattice artefacts
that (assuming $\mu \gg \Lambda$ )
\begin{eqnarray}
   {\oO}(\mu) - {\oO}_{\rm lat}(a,\mu) 
              \simas{a\mu \to 0} a^2 \mu^2 \,.
              \label{eq:sym}
\end{eqnarray}
Here, the challenge is that one wants $\mu$ to be high such
that $\alpha_\msbar(\mu)$ is small and the expansion 
\eq{eq:alpha_series} is precise and
at the same time $a\mu$ small because of \eq{eq:sym}.
Recalling \eq{eq:largev} one is usually in the blue shaded
region of \fig{fig:alpha_lattice_landscape} and
it is difficult to extrapolate when $\alpha_\msbar$ is small, say 
$\alpha_\msbar \leq 0.3$. 
One has to compromise between the two
requirements. 
At this conference, Peter Petreczky reported about recent progress using this method.

\paragraph{(II) Lattice observables at the cutoff.}
There is also the possibility to consider 
lattice observables involving distances of a few lattice spacings,
which are not related to a continuum observable. The prominent examples
are rectangular Wilson loops $W(r,t)$ of  extent $r \times t$ with 
$r=a m$ and $t=a n$, keeping the integers $n,m$ fixed as one takes the 
limit $a\to0$; the loops shrink to size zero in the limit.
Still such observables have an expansion 
\begin{eqnarray}
W(na,ma) &\simas{g_0\to 0}& \sum_{k\geq0} c_{m,n}^{(k)} \, g_0^{2k} 
    \;\simas{a\to 0} \; \sum_{k\geq0} \hat c_{m,n}^{(k)}\, \gbar_\msbar^{2k}(1/a) \,,
    \label{eq:wmn}
\end{eqnarray}
where in the second step use is made of the relation between
the bare coupling and a renormalized coupling at the cutoff scale,
$g_0^{2} =  \gbar_\msbar^{2}(1/a) + \rmO(\gbar^4)$. 

We note two features of this approach: {\it (i)} Lattice perturbation theory 
has to be used and there is less experience with the size of 
higher-order terms (even after tadpole improvement \cite{Lepage:1992xa}) and the available loop orders are often lower than for continuum perturbation theory. 
{\it (ii)} Lattice artefacts can only be separated from perturbative
corrections in \eq{eq:wmn} by assuming some functional form and fitting to it. 

\paragraph{(III) Continuum-limit observables in small volume and step scaling.}
Early on, there was the idea~\cite{Luscher:1991wu} of using the freedom in the definition of
${\oO}(\mu)$ to consider finite volume quantities, e.g. in $L^4$ 
geometries. Then the renormalization scale in  $\as(\mu)$
is 
\begin{eqnarray}
\mu =1/L \,,
    \label{eq:finitevol}
\end{eqnarray}
and any other dimensionful parameter present in the definition of 
${\oO}(\mu)$ is in a fixed relation to (scaled with) $L$. The advantage is that now
$\mu a = a/L$ can easily be taken to
$a/L=1/8 \ldots 1/32$ or smaller. However,
a number of steps are needed to connect 
recursively 
\begin{eqnarray}
\mu_0 \to s\mu_0 \to s^2 \mu_0 \to \ldots \to s^N \mu_0  \,,
    \label{eq:steps}
\end{eqnarray}
and in each step a few different lattice spacings $a$ have to be simulated to
take the continuum limit. 
The start value $\mu_0$ has to be taken such that the lattice spacings are known in 
$\fm$ through a large volume scale setting sketched above. Thus one typically
starts with $\mu_0 = \rmO(1 \fm^{-1}) = \rmO(200\MeV)$ and needs e.g. $s=2, \, N=9$ in
order to be beyond the $Z$ mass, where perturbation theory can be applied with great confidence.
Indeed, at significantly lower energies, 
one has to be more careful, as seen 
explicitly by the investigation in~\cite{DallaBrida:2018rfy}.
There a family 
of schemes all with a well-behaved three-loop $\beta$-function were studied non-perturbatively.
Depending on the exact choice of scheme, truncation errors of perturbation theory  were found to be large at the level of precision that we are discussing for 
$\alpha_\msbar$.
At this conference, Alberto Ramos presented a recent precise three-flavor computation using this method. For details we refer to his contribution.

\subsection{Towards the 2019 FLAG review.}
The Flavour Lattice Averaging Group (FLAG) formed a working group (R. Horsley, T. Onogi, R.S.) on $\alpha_s$ in 2011 and 
first included 
determinations of $\alpha_s(m_Z)$ in its review 
in 2013~\cite{Aoki:2013ldr}. 
An update appeared in 2016~\cite{Aoki:2016frl}. Presently we are
preparing the 2019 edition. Here I will report on its
{\bf preliminary} status and give a preliminary world average. However, first I will roughly explain the algorithm to arrive
at averages. It has similarities to PDG but still differs 
significantly from the PDG practice. 
The main difference is that FLAG formulates a 
set of  criteria, which computations have to pass 
in order to enter the average of a given quantity of 
phenomenological interest~\cite{Aoki:2016frl}. 

\paragraph{Criteria.}
We cite here from the present draft of the new FLAG review:
\\
{\it 
The major sources of systematic error are common to most lattice
calculations. These include, as discussed in detail below,
the chiral, continuum and infinite-volume extrapolations.
To each such source of error for which
systematic improvement is possible we
assign one of three coloured symbols: green
star, unfilled green circle
\ldots or red square.
These correspond to the following ratings: 
\begin{itemize}
\setlength{\itemsep}{-2pt}
\item[\good]  the parameter values and ranges used 
to generate the datasets allow for a satisfactory control of the systematic uncertainties;
\item[\soso] the parameter values and ranges used to generate
the datasets allow for a reasonable attempt at estimating systematic uncertainties, which
however could be improved;
\item[\bad] the parameter values and ranges used to generate
the datasets are unlikely to allow for a reasonable control of systematic uncertainties.
\end{itemize}
The appearance of a red tag, even in a
single source of systematic error of a given lattice result,
disqualifies it from inclusion in the global average.
}

The last sentence is a first difference to the PDG procedure.

For the computations of $\alpha_s$, the criteria for 
chiral and infinite volume extrapolations are relaxed
as they do not play a dominant role. Instead criteria
on {\it perturbative behaviour} and {\it renormalization scale}
try to make sure that the computation is at reasonable high $\mu$,
the perturbative knowledge is sufficiently good (i.e., $n_\mathrm{l}$, 
the number of loops, is sufficiently high) 
and $\mu$ could be varied over some range in order to confirm 
the perturbative $\mu$-dependence. 
We do not have the space here to discuss this in detail, 
but the general idea is that these criteria try to make
sure that the available Monte Carlo data have a few points 
located sufficiently 
low in the landscape of \fig{fig:alpha_lattice_landscape}, while the 
continuum limit criterion requires to not be too far on the right.
Details are listed in \cite{Aoki:2016frl}; changes in 
FLAG 2019 will be minor.

\paragraph{$\boldsymbol{\as}$ from different methods.}
We now discuss the status of the results for 
$\alpha_{\overline{\rm MS}}^{(5)}(m_Z)$.  In Tab.~\ref{tab_alphmsbar18}
all relevant computations are listed. 
They are organized according to the different methods, namely the different 
observables $\oO$ used. The last column lists the loop order defined above.
We go through the different methods, following the 
classification (I-III) of before.

\paragraph{(I) Continuum-limit observables in large volume.}
There is a large number of different methods. They share the necessity for finding
a compromise between large $\mu$ and small $a \mu$. In the cases where 
computations qualify for taking an average (i.e., there is no red square), 
we perform a weighted average of the 
different results. This yields the mean of the quoted pre-ranges in 
Tab.~\ref{tab_alphmsbar18}. 
According to our judgement the uncertainties are dominantly systematic. 
They are due to the truncation error
of perturbation theory, whether ordinary higher order 
or non-perturbative effects. The question is always whether the values of $\mu$ 
are high enough. We just estimate the perturbative truncation error 
and take this
as the uncertainty of the pre-range. Comparing with the errors of the individual
collaborations one sees that we are somewhat more conservative in our estimate
of the perturbative uncertainty, which seems a good strategy if one wants to arrive
at a safe final range. The individual methods are
(we partially have to simplify)
\begin{itemize}
\setlength{\itemsep}{-2pt}
\item[1] $Q$-$\bar{Q}$ potential:  
    $\quad \oO(\mu) = r^2 F_\mathrm{static}(r)\,,\quad \mu=2/r$, 
    where $F_\mathrm{static}(r)$ is the force between static quarks
    defined by the large-$t$ behaviour of Wilson loops $W(r,t)$. 
    Note that $\nl$ is 3 but $\nl>3$ terms proportional to  $\log\alpha$ are also known. 
     Indeed, at fixed order perturbation theory, starting from three loops,
    there are infrared divergences. As mentioned in Sect.~\ref{sec:static} 
    and references cited there,
    these divergences cancel once contributions from the scale $\alpha_s/r$ are resummed, leaving terms such as 
    $\alpha^4 \log\alpha$ in $\oO(\mu)$, which, in turn, can be resummed to all orders via renormalization group equations. 
\item[2] vacuum polarization:  $\quad \oO(\mu) = D(Q^2)\,,\quad \mu^2=Q^2$,
    with $D$ the Adler function derived from 
    $\Pi^{\mu \nu}_{ij,V+A}(q)$ (\eq{eq:pi_v2}) at Euclidean $q$ and $i\ne j$.
    This method does not yet enter the average 
    as the presently best rating is (\soso\good\bad) \cite{Hudspith:2018zlq}. 
\item[3] Hl current, two points: moments of heavy-light pseudoscalar-current two-point functions. Heavy quarks of masses around the charm and heavier are used. This method has attracted a lot of attention. Different discretizations are used that  allow also to 
compare the continuum-limit moments before the extraction of $\as$. There is quite good agreement, but some tensions exist.
\item[4] gluon-ghost vertex: using gauge fixing, the momentum-space vertex is used. 
This method does not yet enter the average as the continuum limit criterion is not passed.
\item[5] Dirac eigenvalues: $\quad \oO(\mu)={\partial \log(\rho(\lambda)) \over \partial \log(\lambda)}
\,,\quad \mu=\lambda$ with $\rho(\lambda)$ the spectral density of the massless 
    Dirac operator. Also this newly introduced method \cite{Nakayama:2018ubk} 
    does not yet pass the continuum-limit criterion.
\end{itemize}
\paragraph{(II) Lattice observables at the cutoff.}
In this category small ($m,n\leq3$) Wilson loops 
$$ 
  \oO(\mu) = W(ma,na)\,,\quad \mu=k/a
$$
and functions thereof (e.g. $\log(W(a,a)$) are used. The scale factor $k$ is 
adjusted to have better apparent convergence of PT. Our estimate of perturbative uncertainties is somewhat bigger than the one of the collaborations.

\paragraph{(III) Continuum-limit observables in small volume and step scaling.}
Finite volume couplings (with Dirichlet boundary conditions in time) 
are used and their $\mu$-dependence is traced to 
$\rmO(100\GeV)$ by step scaling. Perturbative errors are negligible and 
statistical errors of the many Monte Carlo computations dominate. In 
\cite{Brida:2016flw} the freedom of
choice for the definition of the coupling is used to actually impose one definition
at energies smaller than 4~GeV and a different one for higher energies. This reduces 
overall uncertainties.
\newcommand{\pp}{\phantom{0}}
\begin{table}[!htb]
   \vspace{3.0cm}
   \footnotesize
   \begin{tabular*}{\textwidth}[l]{l@{\extracolsep{\fill}}rlllllllr}
   Collaboration & Ref. & $N_f$ &
   \hspace{0.15cm}\begin{rotate}{60}{publication status}\end{rotate}
                                                    \hspace{-0.15cm} &
   \hspace{0.15cm}\begin{rotate}{60}{renormalization scale}\end{rotate}
                                                    \hspace{-0.15cm} &
   \hspace{0.15cm}\begin{rotate}{60}{perturbative behaviour}\end{rotate}
                                                    \hspace{-0.15cm} &
   \hspace{0.15cm}\begin{rotate}{60}{continuum extrapolation}\end{rotate}
      \hspace{-0.25cm} & 
       $\alpha_\msbar(M_\mathrm{Z})$ & method  & $n_\mathrm{l}$ \\
   & & & & & & & & \\[-0.1cm]
   \hline
   \hline
   & & & & & & & & \\[-0.1cm]
   {ALPHA 17}
            & \cite{Bruno:2017gxd}    & 2+1       & \gA 
            & \good   & \good    & \good 
            & $0.11852(\pp84)$
            & step scaling
            & 2                                   \\
  PACS-CS 09A& \cite{Aoki:2009tf} & 2+1 
            & \gA &\good &\good &\soso
            & $0.11800(300)$
            & 
            & 2                                        \\[1ex]
  \multicolumn{3}{l}{pre-range (average)}  & & & & & 0.11848(\pp81)             &      
  \\[1ex] \hline & & & & & & & & \\[-0.1cm]
   {Takaura 18}
            & \cite{Takaura:2018lpw,Takaura:2018vcy} & 2+1  & \oP 
            & \bad  & \soso  & \soso
            & $0.11790(70)(^{+130}_{-120})$
            & $Q$-$\bar{Q}$ potential
            & 3                            \\[-1ex]
            &&&&&&&             & = static energy (sect.~\ref{sec:static}) \\[-1ex]

   {Bazavov 14}
            & \cite{Bazavov:2014soa}    & 2+1       & \gA & \soso
            & \good   & \soso
            & $0.11660(^{+120}_{-80})$
            & 
            & 3                            \\[1ex]
   {Bazavov 12}
            & \cite{Bazavov:2012ka}   & 2+1       & \gA & \soso
            & \soso  & \soso
            & $0.11560(^{+210}_{-220})$ 
            & 
            & 3                            \\[1ex]
  \multicolumn{5}{l}{pre-range with estimated pert. error}    & & & 0.11660(160)  &      
&      
  \\[1ex] \hline & & & & & & & & \\[-0.1cm]
    Hudspith 18 
            & \cite{Hudspith:2018bpz}    & 2+1       & P
            & \soso  & \soso     & \bad
            & $0.11810(270)(^{\pp+80}_{-220})$
            & vacuum polarization
            & 3      \\      
   JLQCD 10 & \cite{Shintani:2010ph} & 2+1 &\gA & \bad 
            & \soso & \bad
            & $0.11180(30)(^{+160}_{-170})$    
            & 
            & 2 
  \\[1ex] \hline & & & & & & & & \\[-0.1cm]

   HPQCD 10& \cite{McNeile:2010ji}& 2+1 & \gA & \soso
            & \good & \good
            & {0.11840(\pp60)}    
            & Wilson loops
            & 2  
            \\
   Maltman 08& \cite{Maltman:2008bx}& 2+1 & \gA & \soso
            & \soso & \good
            & {$0.11920(110)$}
            & 
            & 2                               \\[1ex]
  \multicolumn{5}{l}{pre-range with estimated pert. error}    & & & 0.11858(120)  &      
&      
  \\[1ex] \hline & & & & & & & & \\[-0.1cm]
  {JLQCD 16}
            & \cite{Nakayama:2016atf}    & 2+1       & \gA 
            & \soso  &  \soso  & \soso
            & $0.11770(260)$
            & Hl current, two points
            & 2                                \\
  {Maezawa 16}
            & \cite{Maezawa:2016vgv}    & 2+1       & \gA 
            & \soso  &  \bad   & \soso 
            & $0.11622(\pp84)$
            & 
            & 2                            \\
       HPQCD 14A 
                    &  \cite{Chakraborty:2014aca} & 2+1+1 & \gA 
                    & \soso & \good   & \soso
                    & 0.11822(\pp74)
                    & 
                    & 2                    \\

   HPQCD 10   & \cite{McNeile:2010ji}  & 2+1       & \gA & \soso
             & \good  & \soso          
             & 0.11830(\pp70)          
             & 
             & 2 \\
   HPQCD 08B  & \cite{Allison:2008xk}  & 2+1       & \gA & \bad
             & \bad  & \bad
             & 0.11740(120) 
             & 
             & 2                                \\[1ex]
  \multicolumn{5}{l}{pre-range with estimated pert. error}    & & & 0.11824(150)  &      
&      
  \\[1ex] \hline & & & & & & & & \\[-0.1cm]
   ETM 13D    &  \cite{Blossier:2013ioa}   & 2+1+1& \gA
                    & \soso & \soso  & \bad 
                    & 0.11960(40)(80)(60)
                    & gluon-ghost vertex
                    & 3                         \\
   ETM 12C    & \cite{Blossier:2012ef}   & 2+1+1 & \gA 
                    & \soso & \soso  & \bad  
                    & 0.12000(140)
		 & 
                    & 3                         \\
   ETM 11D   & \cite{Blossier:2011tf}   & 2+1+1 & \gA 
             & \soso & \soso & \bad  
                    & $0.11980(90)(50)(^{\pp+0}_{-50})$
                    & 
                    & 3                         
\\[1ex] \hline & & & & & & & & \\[-0.1cm]
  
  {Nakayama 18}
            & \cite{Nakayama:2018ubk}    & 2+1       & \gA
            &     \good  &  \soso      & \bad
            & $0.12260(360)$
            & Dirac eigenvalues
            & 2                             \\[1ex]
   & & & & & & & & \\[-0.1cm]
   \hline
   \hline
\end{tabular*}
\begin{tabular*}{\textwidth}[l]{l@{\extracolsep{\fill}}lllllll}
\multicolumn{8}{l}{\vbox{\begin{flushleft} 
\end{flushleft}}}
\end{tabular*}
\vspace{-0.3cm}
\caption{Results for $\alpha_\msbar(M_\mathrm{Z})$ from simulations that use $2+1$ 
or $2+1+1$ flavours of quarks.
A weighted average of the pre-ranges
gives $0.11823(57)$, using the smallest pre-range gives $0.11823(81)$
and the average size of ranges  as an error gives $0.11823(128)$.
         }
\label{tab_alphmsbar18}
\end{table}

\paragraph{World average from FLAG.}
\begin{figure}[!htb]
   \begin{center}
      \includegraphics[width=.8\linewidth]{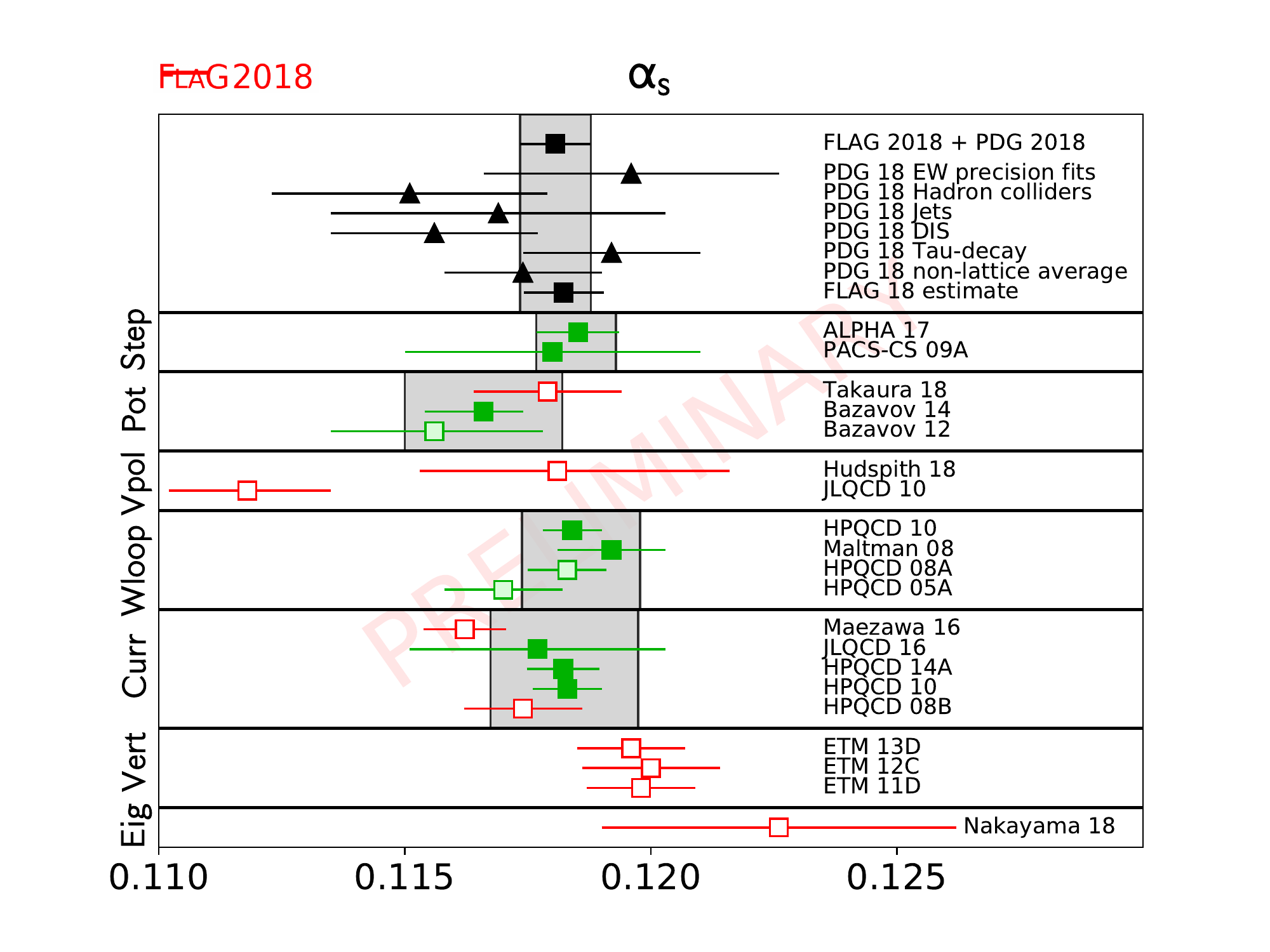}
   \end{center}
\caption{$\alpha_{\overline{\rm MS}}^{(5)}(m_Z)$, the coupling
  constant in the $\overline{\rm MS}$ scheme at the $Z$ mass. 
  The PDG 18 entries give the outcome of their analysis
  from various phenomenological categories including their average.
  The lattice computations with a filled green box symbol have no
  red box in the previous ratings  and therefore qualify for
  averaging. An open green square means the same 
  but the number does not enter an average because it is superseded 
  by a later more complete computation or it was not published at the
  September 2018 deadline. Computations with open red squares do not enter 
  the averages because they had at least one red square before.
  }
\label{alphasMSbarZ}
\end{figure}

The entries of \tab{tab_alphmsbar18} are also displayed in \fig{alphasMSbarZ}.
For each method, the gray band shows the pre-average as explained above. 
We are left with the task to combine those pre-averages. Again we take the central value 
from their weighted average. However, since the errors of the pre-averages are mostly systematic, 
we feel that the straight error $0.00057$ of the weighted average is too optimistic
-- it would be correct for independent Gaussian distributions.
Instead we use the smallest error of the pre-averages.
This yields the {\bf preliminary} result
\begin{eqnarray}\label{eq:FLAGaver}
  \mbox{preliminary: } \alpha_{\overline{\rm MS}}^{(5)}(m_Z) = 0.11823(81)\qquad\Refs~\mbox{\cite{Bruno:2017gxd,Nakayama:2016atf,Bazavov:2014soa,Chakraborty:2014aca,McNeile:2010ji,Aoki:2009tf,Maltman:2008bx}}, 
\end{eqnarray}
and the associated $\Lambda$ parameter
\begin{eqnarray}
   \mbox{preliminary: }\Lambda_{\overline{\rm MS}}^{(5)} = 211(10)\,\MeV\hspace{5mm}\qquad\Refs~\mbox{\cite{Bruno:2017gxd,Nakayama:2016atf,Bazavov:2014soa,Chakraborty:2014aca,McNeile:2010ji,Aoki:2009tf,Maltman:2008bx}}.
\end{eqnarray} 
The PDG has unfortunately not updated its world average since its
2016 value
\begin{eqnarray}
   \alpha^{(5)}_{\overline{\rm MS}}(m_Z) &=& 0.1174(16) \,, \quad 
   \mbox{PDG 2016, non-lattice \cite{Patrignani:2016xqp}} \,.
\end{eqnarray}
That number originates from the different pre-averages listed in
the upper box of \fig{alphasMSbarZ}. We can now add the information from the lattice
and arrive at an average 
\begin{eqnarray}
    \mbox{preliminary: }\alpha^{(5)}_{\overline{\rm MS}}(m_Z) &=& 0.11806(72) \,, \quad 
   \mbox{FLAG 2018 + PDG 2018/2016  }
\label{PDG_FLAG_alpha}  
\end{eqnarray}
of PDG non-lattice and FLAG lattice (weighted average).
The error is reduced significantly compared to FLAG 2016 and PDG 2016 and almost as small as PDG 2014. 
Unfortunately \eq{PDG_FLAG_alpha} does not yet contain the interesting new non-lattice analyses discussed at this conference.

\subsection{Further progress}

Here I collect some lessons which I have learned
during the period when I was involved in forming and 
discussing a world average for $\as$.

The basic problem is simple and has been spelled out often,
phrased in varying words. In order to have a 
precise value with an error that can be estimated  
by perturbation theory itself,
 large energy scales $\mu$ have to be reached and
 theory assumptions have to be kept at a minimum.
I think that we will not make further progress if we include 
complicated processes, where non-perturbative contributions
have to be fitted or removed by complicated analyses in order
to make lower energies accessible.
Dealing with non-perturbative physics is always based on 
assumptions -- if only where the expansion in $1/\mu$ applies
and lowest-order terms $(1/\mu)^{N_\mathrm{min}}$ dominate.  

We should therefore  separate the  determination of $\as$
at high enough $\mu$, simple theory, from 
tests of perturbation theory, with resummations, studies of 
higher-twist contributions, etc. 

The concept of  criteria introduced by FLAG is very useful
in this respect and I would advocate to consider such a procedure
for phenomenological determinations.
One should at least consider a
 criterion on minimum values of $\mu$, paired with
sufficiently high perturbative order. In FLAG these are the
``renormalization scale'' / ``perturbative behaviour'' criteria.

I personally also think that the  criteria of FLAG need
to be made more strict as time goes on. This is necessary to avoid 
situations where complicated procedures, involving e.g. separate
estimates of perturbative errors (see above) are needed to arrive at a safe
range. 

Finally, it seems that the limit of lattice determinations 
of $\as$ is not yet reached; I believe a factor two reduction in the error is 
possible with some variation of the developed techniques and some dedication.
A rather tough limitation, which is beyond such a factor of two, 
may be the inclusion of electromagnetic effects.
They dominantly (by far) enter in the process of scale setting.
At present one essentially uses models for relating Nature at low energy
to pure QCD with electromagnetic interactions removed. A typical estimate
of the precision is at the level significantly below a \% for, e.g., $f_\pi$,
defined in pure QCD related to $\pi \to \ell \nu + \gamma$'s
decays in Nature \cite{Patrignani:2016xqp}.
Uncertainties due to that step can be translated directly into 
those of the 3-flavour $\Lambda$-parameter which is at the level 
of 4\% right now. In other words, there is still some room before 
that limit becomes relevant.

\section{Final discussion}
\label{sec:finaldiscussion}

The spectacular progress achieved in higher-order QCD calculations has made possible to predict many observables with an impressive NNLO accuracy, reaching even the N${}^3$LO for fully inclusive quantities such as $R_Z$ and $R_{\tau,V+A}$, as well as for the static energy. 
With this improved theoretical control, rather clean and precise determinations of the strong coupling have come out in a rich variety of energy scales. At the same time, high-precision lattice computations have been able to control the 
continuum limit in $\as$ determinations with NNLO accuracy up to energies around the Z mass.
The addition of novel LHC processes directly sensitive to $\as$,
and for which NNLO QCD calculations are available, have allowed
several recent determinations from global PDF fits with small
uncertainties; being able to estimate the theory errors associated to these
determinations arising from MHO
is now essential to make further progress in this direction.

The most precise determinations of $\alpha_s(m_Z)$ are currently obtained from lattice simulations. The FLAG lattice average in Eq.~(\ref{eq:FLAGaver}) clearly dominates any world average with non-lattice results, making manifest the importance of having a strict and reliable control of lattice and 
perturbation theory systematics;  before averaging the results,
FLAG tries to ensure their quality by a set of criteria.
The good agreement between the FLAG average and the accurate non-lattice  determinations presented in Eqs.~(\ref{eq:alpha_Z}), (\ref{eq:alpha-tauMZ}) and (\ref{eq:NNPDFalpha}) constitutes a highly non-trivial consistency test 
among results obtained with quite different techniques and physical observables. Still, there remain some 
results which are quite a bit lower.
In this report we mentioned Eqs.~\eqref{eq:thrust} and \eqref{eq:C-par}, as well as the ABMP16 result in Fig.~\ref{fig:alphas_pdffit}.
These differences need to be better understood.

A precise determination of $\alpha_s(m_Z)$ is of paramount relevance, since it fixes the unique coupling of QCD and, therefore, its predictions for any physical system.
Nevertheless, we must also emphasize the very important added value of precisely measuring the strong coupling at different scales. The accurate low-energy determinations from $\tau$ decay, in Eq.~(\ref{eq:alpha-tau}), and from a number of lattice determinations such as the QCD static energy,  in Eq.~(\ref{eq:StaticEalpha}), provide the needed inputs to perform a very significant test of the running of $\alpha_s$. The beautiful agreement with measurements performed at much higher energies gives a fundamental verification of the asymptotic-freedom property of QCD.
In addition, LHC cross sections have also the unique capability of
directly measuring
the running of $\alpha_s(Q)$ far above the electroweak scale,
providing a unique test of the Standard Model and of new
strongly interacting sectors at high energies.

\acknowledgments
We thank the organizers of the ``XIIIth Quark Confinement and the Hadron
  Spectrum'' conference for inviting us to participate
in this round table about the strong coupling constant.
R.~S. thanks Roger Horsley and Tetsuya Onogi as well as the other members 
of FLAG for a fruitful collaboration and many vivid discussions, and
Roger Horsley for preparing \fig{alphasMSbarZ} as well as input on an
early version of this manuscript.
A.~V. thanks Johannes Weber for providing Fig.~\ref{fig2}.
This work has been supported by the DFG and the NSFC through funds provided to the Sino-German CRC 110 ``Symmetries and the Emergence of Structure in QCD'', 
by the DFG cluster of excellence ``Origin and structure of the universe'',
the ERC from the Starting Grant ``PDF4BSM'', the Dutch National
Organisation for Scientific Research,
by  the  Spanish  State  Research Agency and ERDF funds from the EU Commission  [Grants  FPA2017-84445-P  and  FPA2014-53631-C2-1-P],   
by  Generalitat   Valenciana   [Grant Prometeo/2017/053] and by the Spanish Centro de Excelencia Severo Ochoa Programme [Grant SEV-2014-0398].

\providecommand{\href}[2]{#2}\begingroup\raggedright\endgroup


\end{document}